\documentclass[lettersize,journal]{IEEEtran}
\usepackage{amsmath,amsfonts}
\usepackage{algorithmic}
\usepackage{algorithm}
\usepackage{array}
\usepackage{subfig}
\usepackage{textcomp}
\usepackage{stfloats}
\usepackage{url}
\usepackage{verbatim}
\usepackage{graphicx}
\usepackage{cite}
\usepackage{amssymb}%大于小于号
\usepackage{cases}%大括号多行编号
\usepackage{svg}%插入图片
\usepackage{multirow}%表格
\usepackage{makecell}
\usepackage{bbm}

\usepackage{booktabs}

\usepackage{hyperref}
\hypersetup{hypertex=true,colorlinks=true,linkcolor=blue,anchorcolor=blue,citecolor=blue}

\newtheorem{corollary}{\bf Corollary}[section]
\newtheorem{remark}{\textit{Remark}}

\newtheorem{lemma}{\bf Lemma}[section]

\newtheorem{theorem}{\bf Theorem}[section]
\newtheorem{definition}{\bf Definition}[section]

\def\MatrixFont{\bf}

\newcommand{\mU}{{\MatrixFont U}}

\begin{document}

\title{BG-FlipIn: A Bayesian game framework for FlipIt-insider models in advanced persistent threats}

\author{Yang Jiao, Guanpu Chen,~\IEEEmembership{Member,~IEEE,}~Yiguang Hong,~\IEEEmembership{Fellow,~IEEE}
        % <-this % stops a space
\thanks{Yang Jiao and Yiguang Hong are with Shanghai Research Institute for Intelligent Autonomous System, Tongji University, Shanghai, 201210, China (e-mail: jy0903@tongji.edu.cn, yghong@iss.ac.cn).}
\thanks{Guanpu Chen is with School of Electrical Engineering and Computer Science, KTH Royal Institute of Technology, Stockholm, 100 44, Sweden (e-mail: guanpu@kth.se).}% <-this % stops a space
\vspace{-10pt}
}

% The paper headers
\markboth{Journal of \LaTeX\ Class Files,~Vol.~14, No.~8, August~2021}%
{Shell \MakeLowercase{\textit{et al.}}: A Sample Article Using IEEEtran.cls for IEEE Journals}

%\IEEEpubid{0000--0000/00\$00.00~\copyright~2021 IEEE}
% Remember, if you use this you must call \IEEEpubidadjcol in the second
% column for its text to clear the IEEEpubid mark.

\maketitle

\begin{abstract}
In this paper, we study advanced persistent threats (APT) with an insider who has different preferences. To address the uncertainty of the insider's preference, we propose the BG-FlipIn: a Bayesian game framework for FlipIt-insider models with an investigation on malicious, inadvertent, or corrupt insiders. We calculate the closed-form Bayesian Nash Equilibrium expression and further obtain three edge cases with deterministic insiders corresponding to their Nash Equilibrium expressions. On this basis, we further discover several phenomena in APT related to the defender's move rate and cost, as well as the insider's preferences. We then provide decision-making guidance for the defender, given different parametric conditions. Two applications validate that our BG-FlipIn framework enables the defender to make decisions consistently, avoiding detecting the insider's concrete preference or adjusting its strategy frequently.

\end{abstract}


\section{Introduction}

\IEEEPARstart{A}{dvanced} persistent threats (APT) have become a major challenge in cybersecurity \cite{alshamrani2019survey}, characterized by long-term, highly sophisticated attacks that target sensitive resources. Approaches to counter APT have gained attention in artificial defense \cite{sun2023impulsive}, reinforcement learning \cite{xiao2024reinforcement,chen2025defending}, and game theory \cite{cheng2022single,yang2018effective,zhang2022game}. Among these approaches, game theory stands out as a powerful framework, as it provides equilibrium-based insights to modeling the strategic interplay between the defender and attacker. Within game-theoretic models, the two-player FlipIt game is a widely adopted approach \cite{van2013flipit}. In FlipIt games, both the attacker and defender can reclaim the control of a shared resource through discrete moves called flips, and there are two typical models: the periodic FlipIt game \cite{chen2017security,pawlick2017strategic} and the exponential FlipIt game \cite{feng2016stealthy}. Nevertheless, most game-theoretic approaches to counter APT, including FlipIt, primarily focus on the bilateral interaction between the defender and attacker.

Notably, insider threats in cybersecurity have garnered increasing attention in recent years. Unlike external attackers, insiders inherently possess privileged access to sensitive resources, which enables them to cause more severe damage to organizational security \cite{brdiczka2012proactive}. Moreover, studies on insiders in cybersecurity have highlighted the importance of diversifying their preferences \cite{sinclair2008preventative,sarkar2010assessing,homoliak2019insight}, like malicious, inadvertent, corrupt, etc. In addition, detecting the certain preference of insiders is usually a challenging task, as it often relies on predefined rules \cite{abdullah2025robust,axelrad2013bayesian}. For example, static models often fail to detect evolving or covert insider behaviors, causing missed detections or false alarms, while changes in insider preferences force rapid adjustments in security policies, potentially undermining employee trust \cite{posey2011understanding}. In fact, there have been a few works focusing on insiders in the scope of APT research \cite{feng2015stealthy,zhang2019mathtt,liu2021flipit,xu2024consistency,liu2020defense}, most of which consider only a single or at most two deterministic preferences of insiders.

On the other hand, the Bayesian game offers a powerful tool for addressing players' uncertain preferences. In this framework, each player is assumed to know the prior probability distribution over the possible types of others. Bayesian games have been widely applied to management \cite{harsanyi1967games} and engineering \cite{akkarajitsakul2011distributed,chen2020learning,zhang2023distributed}, because this approach enables the modeling of strategic interactions under uncertainty, especially for incomplete information. Unsurprisingly, the Bayesian game has already been adopted in APT research to capture the interaction between attacker and defender \cite{huang2020dynamic,halabi2021protecting}. To the best of our knowledge, no prior work employed the Bayesian game to characterize the insider's preferences in APT.

In this paper, we are motivated to address the FlipIt-insider challenge in APT where the insider has uncertain preferences. To this end, we propose the BG-FlipIn: a Bayesian game framework for FlipIt-insider models who investigates malicious, inadvertent, and corrupt insiders. This unified framework enables the defender to make decisions while reducing the cost of detecting insider’s preference. Also, it provides a consistent defense strategy to avoid frequent adjustments especially when the insider's preference switches rapidly.

The main contributions are summarized as follows:
\begin{itemize}
\item We propose a Bayesian game framework to address the FlipIt-insider challenge in APT problems, where the insider has uncertain preferences. Unlike existing models focusing merely on deterministic insiders, our framework provides a consistent defense strategy in APT by enhancing the defender’s decision-making capability under potential uncertainties.

\item We perform a rigorous Bayesian Nash Equilibrium (BNE) analysis by calculating the closed-form expression in different parametric conditions. We also consider three edge cases where the deterministic insider is malicious, inadvertent, or corrupt, and reveal their corresponding Nash Equilibrium (NE) as well. All the equilibrium expressions show a clear dependency on system parameters and the insider preferences.

\item Based on the equilibrium results, we discover some significant phenomena in APT related to the defender's move rate and cost, together with
the insider's preferences. We then provide decision-making guidance for the defender given different parametric conditions. Moreover, we identify a parameter interval where BNE outperforms all edge-case NEs, reflecting the advantage of the Bayesian framework.

\item  We further present two applications with intuitive evidence in their illustrations to validate our theoretical results. One simulation is conducted under unknown insider preferences to show the effectiveness of our framework, while the other experiment is carried out in cloud-enabled remote state estimation to evaluate our approach when the insider rapidly changes its preferences.  
% These applications demonstrate our BG-FlipIn framework’s effectiveness in coping with uncertain insider preferences.
\end{itemize}

The rest of the paper is organized as follows: Section II provides a literature review. Section III revisits the FlipIt game and classifies insider preferences. Section IV introduces the Bayesian game for FlipIt-insider models, considers three edge deterministic cases, and analyzes the corresponding BNE and NEs. Section V reveals phenomena shared with existing research and those unique to our framework, provides decision-making guidance for the defender, and analyzes the advantage of the Bayesian framework. Section VI presents two applications. Finally, Section VII concludes the paper.

\section{Related work}

In this section, we provide a literature review on the topics in this paper.

% \subsection{FlipIt game in APT}

\textbf{FlipIt game in APT:} Originally proposed by van Dijk et al. \cite{van2013flipit}, the FlipIt game models a two-player competition for the control of a shared resource. Within this framework, strategies for determining flip intervals are typically categorized into non-adaptive and adaptive classes. Two classical non-adaptive models are widely employed in APT. The first is the periodic FlipIt game where a player flips at a fixed interval. Based on this model, a contract-based FlipIt game is developed in \cite{chen2017security} to assess security risks and cloud quality of service under APT. A signaling game is combined in \cite{pawlick2017strategic} with the periodic FlipIt game to model strategic trust in cloud-enabled cyber-physical systems (CPS) under APT. Another model is the exponential FlipIt game in which flips follow a Poisson process. In this context, an exponential defense strategy is considered in \cite{feng2016stealthy} to prevent the APT attacker from exploiting feedback. The periodic FlipIt game is adopted as the basic model in this work, as it captures regular defensive checks or persistent attacks and yields linear benefit functions.

\textbf{Insiders in cybersecurity:} Insiders play a critical role in cybersecurity. The preferences of insiders are typically categorized as malicious, inadvertent, corrupt, etc. For example, a malicious insider may access sensitive data without authorization \cite{sarkar2010assessing}; an inadvertent insider may be an employee who falls victim to a phishing attack \cite{homoliak2019insight}; and a corrupt insider may betray their organization for personal profit \cite{sinclair2008preventative}. On the other hand, game-theoretic approaches are widely employed to capture insider threats. The interactions among the defender, attacker, and insider are formulated as a three-player leader–follower game in \cite{xu2024consistency} to analyze the consistency between the Stackelberg equilibrium and NE. A security resource allocation game is developed in \cite{liu2020defense}, in which an insider may probabilistically leak the protection status of certain measurements. 

Some studies further extend the FlipIt game to address insider threats in APT. A FlipIt-insider game is proposed in \cite{feng2015stealthy} in which a corrupt insider can trade information to the attacker for profit. In addition, a FlipIt model with cyber insurance is developed in \cite{zhang2019mathtt}, where the insider acts as the insurer. Another research integrated a semi-Markov process with the FlipIt game to model cyber attacks, considering malicious insider assistance \cite{liu2021flipit}. Nonetheless, all of these models focus on a certain insider preference and cannot address situations where the insider’s preference is uncertain.

\textbf{Bayesian game in CPS:} The Bayesian game has been widely applied in CPS scenarios due to its ability to model interactions under incomplete information. For instance, learning the inherent attackers in repeated Bayesian network games is addressed in \cite{chen2020learning}. Abstracted from electricity markets, the subnetwork zero-sum game problem and its BNE are studied in \cite{zhang2023distributed}. Also, a Bayesian game is explored in \cite{bhatia2020game} to develop a computing platform for quantifying the probability of food quality. Trust management for agricultural green supply is designed in \cite{bai2021blockchain}, where a Bayesian game ensures the data reliability provided by different sensors. 

Although Bayesian game models have been employed in APT, most studies focus on defender–attacker interactions and do not incorporate insiders. For instance, a multi-stage Bayesian game framework is proposed in \cite{huang2020dynamic} for proactive defense against APT. A Bayesian Stackelberg game is designed in \cite{halabi2021protecting} to defend against APT in the Internet of Vehicles. Notable research gaps remain in addressing the potential uncertainty in insider preferences, which motivated us to construct the BG-FlipIn in this work.

\section{Preliminary}

In this section, we introduce the fundamental concepts underlying our work, including the periodic FlipIt game and insider preferences. 

\subsection{Revisiting FlipIt game}

In the two-player FlipIt game, both the defender and the attacker can reclaim control of a shared resource by a move called a ‘flip’, which alternates ownership between them with each move. The following are the main rules: 
\begin{itemize}
\item Time is continuous and infinite.

\item The player is unaware of the period during which the opponent has taken control of the resource, as well as the current ownership of the resource, unless they make a move themselves.

\item The resource in the FlipIt game is a whole entity and cannot be partially controlled.

\item Players earn rewards by controlling the resource and aim to maximize their control time.
\end{itemize}

\begin{figure}[t]
    \centering
    \includegraphics[width=1.0\columnwidth]{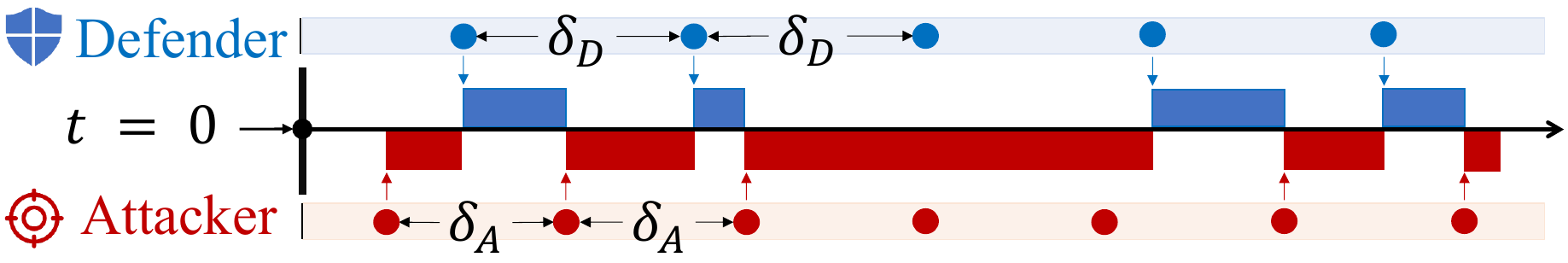}
    \caption{The periodic FlipIt game model between the defender and attacker}
    \label{fig1}
\end{figure}

We adopt the periodic FlipIt game as the basic model, as it captures regular defensive checks or persistent attacks and yields linear benefit functions \cite{feng2015stealthy,liu2021flipit,zhang2019mathtt}. In this FlipIt game, both players employ a periodic strategy with random phase. This strategy involves the player moving with fixed interval $\delta$ and choosing the time of the first move uniformly at random in interval $[0, \delta]$. Specifically, let $\alpha$ and $\beta$ represent the average move rate of the defender and the attacker, respectively. Furthermore, let $\delta_D = \frac{1}{\alpha}$ and $\delta_A = \frac{1}{\beta}$ be their respective periods. Then the periodic FlipIt game model can be illustrated in Fig. \ref{fig1}.

Let $x$ denote the average time when the resource is protected by the defender. According to \cite{van2013flipit}, we can consider two cases to compute $x$: If $\alpha \leqslant \beta$, then $\delta_A \leqslant \delta_D$. In an attacker's move interval, the probability that the defender moves is $\frac{\alpha}{\beta}$. Moreover, the defender has at most one move in this interval because $\delta_A \leqslant \delta_D$ and their move is uniformly distributed at random in $[0,\delta_D]$. Therefore, the expected period of attacker control within the interval is $\frac{\alpha}{2\beta}$. Similarly, if $\alpha > \beta$, we obtain $x=1-\frac{\beta}{2\alpha}$ in the same manner.

Let $C_D$ and $C_A$ represent the move cost for the defender and the attacker. The benefit function can be expressed as:
\begin{equation}
\label{flipit_benefit_function}
\left\{
    \begin{aligned}
        &\mU_D=x -  C_D\alpha,\\
        &\mU_A=1-x-  C_A\beta.
    \end{aligned}\right.
\end{equation}
Then the classic periodic FlipIt game can be written as:
\begin{equation}
\label{G_0}
    G=\left\langle \mathcal{I}, (S_i), (\mU_i) \right\rangle,
\end{equation}
where $\mathcal{I}=\{D, A\}$, $S_D=[ 0, \alpha_m]$, $S_A=[ 0, \beta_m]$, and the benefit function $\mU_D$, $\mU_A$ are defined in (\ref{flipit_benefit_function}). Here $\alpha_m$ and $\beta_m$ are sufficiently large constants.

\subsection{Insider preferences}

An insider is an individual with legitimate and privileged access to an organization’s internal resources \cite{sinclair2008preventative,sarkar2010assessing}. There are a variety of insider preferences \cite{sinclair2008preventative}, for example, malicious, inadvertent, selfish, etc. We investigate insider preferences with the following three broad categories:

-\textbf{Malicious insider:} who deliberately intends to steal the defender's resource, often for financial gain or personal revenge \cite{glancy2020classification}. Since revenge is irrational and difficult to model using game theory, we focus on those seeking to benefit from stealing confidential information \cite{bushman2001time}. In APT attacks, malicious insiders exploit organizational trust and steal sensitive data over extended periods \cite{willison2013beyond}.

-\textbf{Inadvertent insider:} who abuses their privileged access to cause resource leakage, without realizing it. They lack harmful intent but can still compromise security \cite{greitzer2014analysis}. In APT scenarios, they may unintentionally assist adversaries by sharing sensitive information, clicking phishing links, or misconfiguring systems \cite{carley2016inadvertent}. 

-\textbf{Corrupt insider:} who prioritizes personal gains over the organization’s interests and lacks collective loyalty. If they are tempted by external interests, they may be bought and betray the organization  \cite{green2023understanding}. In APT attacks, corrupt insiders may contact attackers proactively or reactively, seeking opportunities to betray their organization for personal profit \cite{shaw2005ten}.

In the periodic FlipIt game (\ref{G_0}), the players compete for control of the resource using periodic strategies. With the introduction of the insider, the game retains its fundamental structure but exhibits the following variations:

\begin{itemize}
    \item The insider does not compete with the defender and attacker for ownership of the resource, and their moves do not change the current ownership of the resource.
    \item The resource in the FlipIt-insider game is no longer a whole entity and can be partially stolen by the malicious and corrupt insiders, or leaked by the inadvertent insider.
    \item We do not consider the impact of the insider on the resource owned by the attacker.
\end{itemize}

\begin{figure}[t]
    \centering
    \includegraphics[width=1.0\linewidth]{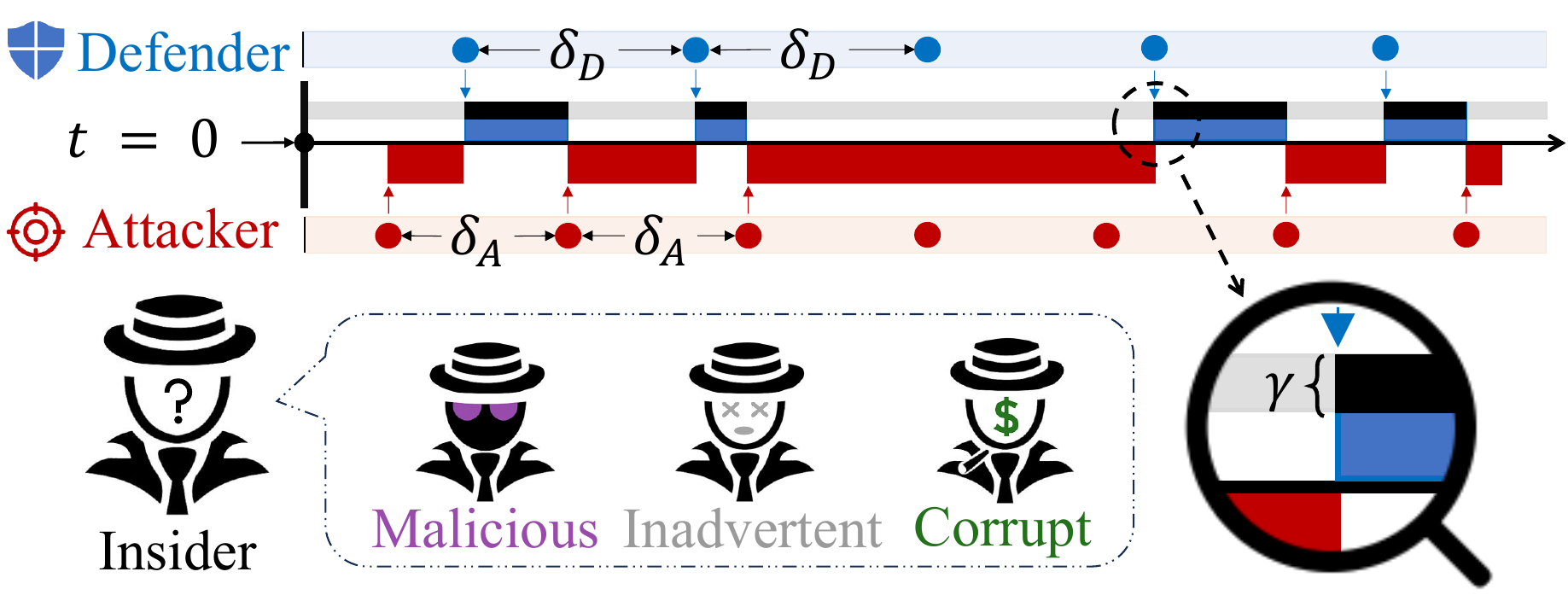}
    \caption{Three preferences of insiders into the periodic FlipIt game (the black part represents the defender’s resources affected by insiders).}
    \label{fig2}
\end{figure}

On this basis, let $\gamma \in [0, \gamma_m]$ represent the percentage of the resource impacted by the insider, where $\gamma_m$ is the upper bound of this proportion. If $\gamma_m = 1$, the insider would cause the defender to lose all control over the resource, leading to the defender exiting the game due to the lack of rewards. If $\gamma_m = 0$, the insider has no impact on the resource, and the game would degenerate into a two-player game. With $0 < \gamma_m < 1$, the FlipIt-insider model is illustrated in Fig. \ref{fig2}, and all subsequent analysis is conducted under this setting.

\section{Bayesian game for FlipIt-insider models}

In this section, we present the BG-FlipIn: a Bayesian game framework for the FlipIt-insider models to capture the uncertainty in insider preferences. We then derive the corresponding BNE, as well as the NEs for three edge cases.

\subsection{The Bayesian game model}

In the previous section, we have classified insider preferences and introduced the insider into the periodic FlipIt game \eqref{G_0}. In APT scenarios, the defender usually faces incomplete information about the insider’s preferences. To this end, we design a Bayesian game for FlipIt-insider models. 

Consider a Bayesian game denoted by 
\begin{equation}
\label{Γ}
    \Gamma=\left\langle \mathcal{I}, (S_i),T, P(\cdot), (f_i)\right\rangle,
\end{equation}
with the set of players $\mathcal{I}=\{D,A,I\}$, and the feasible strategy set $S_D=[ 0, \alpha_m]$, $S_A=[ 0, \beta_m]$, $S_I=[ 0, \gamma_m]$. For each player $i$ in the set $\mathcal{I}$, the incomplete information is referred to as their types, denoted as $t_i \in T=\{t_1,t_2,t_3\}$. Specifically, for all $i \in \mathcal{I}$, $t_i = t_1$ denotes that player $i$ is engaged in the FlipIt game with a malicious insider, $t_i = t_2$ denotes engagement in the FlipIt game with an inadvertent insider, and $t_i = t_3$ denotes engagement in the FlipIt game with a corrupt insider. The tuple $\boldsymbol{t} = (t_D, t_A, t_I) \in \boldsymbol{T}$ represents a random variable that maps from the probability space $(\Omega, \mathcal{B}, P)$ to $\mathbb{R}^3$. The density function of $P(\cdot)$ is denoted by $p$, with the marginal density defined as $p_i(t_i) = \sum_{t_{-i}\in T_{-i}} p(t_i, t_{-i})$ and the conditional probability density given by $p_i(t_{-i} \mid t_i) = p(t_i, t_{-i}) / p_i(t_i)$ for $i \in \mathcal{I}$. Here, each player $i \in \mathcal{I}$ only knows its own type but not those of its rivals, but the joint distribution $P$ is publicly accessible. The benefit function for player $i$ is formulated as $f_i: S_D \times S_A \times S_I \times T \rightarrow \mathbb{R}$, depending on all players' strategies and the type of player $i$.

Then the conditional expected benefit of a player of type $t_i$, $i \in \mathcal{I}$, playing strategy $\alpha \in S_D$, $\beta \in S_A$ or $\gamma \in S_I$, is
\begin{equation}
\label{u_i}
U_i(\alpha,\beta,\gamma,t_i)=\sum\nolimits_{t_{-i}\in T_{-i}}f_i(\alpha,\beta,\gamma,t_i)p_i(t_{-i} \mid t_i).
\end{equation}
The expected benefit of each player can be obtained as
\begin{equation}
\label{expect}
\tilde{\mU}_i(\alpha,\beta,\gamma)=\mathbb{E}\big(U_i(\alpha,\beta,\gamma,T)\big), \forall i \in \mathcal{I}.
\end{equation}

Now we define the benefit functions for all players.

\textbf{Insider:} The insider’s benefit depends on its preference, with $C_I$ representing the unit cost incurred by the insider for influencing a percentage of the defender's resource.

The malicious insider aims to undermine the defender’s control by stealing resources, while bearing a move cost. The benefit function for the malicious insider is given by
\begin{equation*}
    f_I(\alpha,\beta,\gamma,t_1)=x\gamma - C_I\gamma,
\end{equation*}
where the first term represents the stolen resource and the last term indicates the cost of the malicious insider when the resource theft percentage is $\gamma$.

The inadvertent insider is unaware that they are involved in a game. Even though their behavior may inadvertently affect the defender's control over the resource, the inadvertent insider does not recognize these consequences and does not need to bear any cost for their moves. Therefore, the benefit function of the inadvertent insider can be characterized as a zero function:
\begin{equation*}
     f_I(\alpha,\beta,\gamma,t_2)=0.
\end{equation*}

The corrupt insider is motivated by the attacker and is indifferent to the potential impact of reduced benefit for the defender on their own benefit. Let $C_{AI}$ be the unit reward given by the attacker for assisting in the corrupt insider's efforts to steal the benefit of the defender, then
\begin{equation*}
 f_I(\alpha,\beta,\gamma,t_3)=-C_I\gamma+C_{AI}\gamma ,
\end{equation*}
where $C_{AI}>C_I$ ensures that the corrupt insider has an incentive to participate. Note that the term $C_I\gamma$ represents the cost incurred by the corrupt insider when exerting effort at level $\gamma$, while the term $C_{AI}\gamma$ denotes the reward provided by the attacker for the same level of effort. This benefit function highlights the corrupt insider's willingness to collaborate with the attacker in exchange for personal gain, while also accounting for the costs associated with their move.

\textbf{Attacker:} The attacker’s benefit generally consists of the expected time controlling the resource minus move costs. When colluding with a corrupt insider, the attacker also bears the additional reward paid to the insider. Therefore, the attacker’s benefit is summarized as  
\begin{equation*}
   f_A(\alpha,\beta,\gamma,t_A)=1-x-C_A\beta-\mathbbm{1}_{\{t_A=t_3\}}C_{AI}\gamma,
\end{equation*}
where $\mathbbm{1}_{\{t_A=t_3\}}$ is an indicator that equals $1$ when the insider is corrupt and $0$ otherwise.

\textbf{Defender:} The defender’s objective is to maximize the protected time of the resource while minimizing move costs. In all three insider scenarios, the defender suffers an additional loss proportional to the resource leakage caused by the insider. Thus, the defender’s benefit is  
\begin{equation}
\label{U_D}
   f_D(\alpha,\beta,\gamma,t_D)=x-C_D\alpha-x\gamma,
\end{equation}
where $x\gamma$ represents the resource stolen or leaked, regardless of whether the insider is malicious, inadvertent, or corrupt.

Subsequently, let $\theta_1$, $\theta_2$, and $1-\theta_1-\theta_2$ be the probability of the insider being the malicious insider, corrupt insider and inadvertent insider, i.e., $p(t_D=t_1,t_A=t_1,t_I=t_1)=\theta_1$, $p(t_D=t_2,t_A=t_2,t_I=t_2)=1-\theta_1-\theta_2$, $p(t_D=t_3,t_A=t_3,t_I=t_3)=\theta_2$, where $\theta_1 > 0$, $\theta_2 > 0$ and $\theta_1+\theta_2 < 1$. Then by (\ref{u_i}) and (\ref{expect}), we obtain \begin{subnumcases}{\label{bayes model}} \tilde{\mU}_D=x - C_D\alpha- x \gamma ,\label{U_D_tilde}\\ \tilde{\mU}_A=1-x-C_A \beta-\theta_2 C_{AI} \gamma, \\ \tilde{\mU}_I=\theta_1(x \gamma-C_I \gamma)+\theta_2 \left(-C_I \gamma+C_{AI} \gamma\right). \end{subnumcases}

Based on the above functions, we define the concept of BNE for the BG-FlipIn as follows.
\begin{definition} Let $\alpha^* \in S_D$, $\beta^* \in S_A$ and $\gamma^* \in S_I$, then the strategy triple $(\alpha^*, \beta^*, \gamma^*)$ is called the BNE of BG-FlipIn if 
\begin{equation*}
\begin{array}{l}
     \tilde{\mU}_D(\alpha^*, \beta^*, \gamma^*) \geqslant \tilde{\mU}_D(\alpha, \beta^*, \gamma^*), \forall \alpha \in S_D, \vspace{1ex}\\
     \tilde{\mU}_A(\alpha^*, \beta^*, \gamma^*) \geqslant \tilde{\mU}_A(\alpha^*, \beta, \gamma^*), \forall \beta \in S_A, \vspace{1ex}\\
     \tilde{\mU}_I(\alpha^*, \beta^*, \gamma^*) \geqslant \tilde{\mU}_I(\alpha^*, \beta^*, \gamma),\forall \gamma \in S_I.
\end{array}
\end{equation*}
\end{definition}

\subsection{Equilibrium of the BG-FlipIn}

In this subsection, we establish the existence and explicit form of the BNE in BG-FlipIn. In addition, we present the NEs for three edge cases when the insider preference is certain.

Based on the benefit functions for all players and their beliefs, we can derive the following theorem, with its proof provided in Appendix \ref{app1}. This theorem presents the existence and explicit form of the BNE for BG-FlipIn. For notation simplicity, we define the attack-defense cost ratio (ADCR) as 
\begin{equation*}
    \sigma=\frac{C_A}{C_D},\quad \sigma \in [0,\infty).
\end{equation*}
\begin{theorem}
\label{theorem_bayes}
If an equilibrium profile $(\alpha^*, \beta^*, \gamma^*)$ is a BNE of BG-FlipIn, then its closed-form expression is subject to the following conditions:
    \begin{itemize}
        \item When $\alpha \leqslant \beta$,
        \begin{itemize}
            \item if $\sigma \leqslant 1$ and $\sigma < (2\theta+2)C_I-2\theta C_{AI}$,
            \begin{equation}
            \label{37}
                (\alpha^*, \beta^*, \gamma^*)=(\frac{C_A}{2{C_D}^2}, \frac{1}{2C_D}, 0),
            \end{equation}
            \item if $ \frac{(2\theta+2)C_I-2\theta C_{AI}}{1-\gamma_m} < \sigma \leqslant \frac{1}{1-\gamma_m}$,
            \begin{equation}
            \label{38}
                (\alpha^*, \beta^*, \gamma^*)=(\frac{C_A (1-\gamma_m)^2}{2{C_D}^2}, \frac{1-\gamma_m}{2C_D}, \gamma_m).
            \end{equation}
        \end{itemize}
        \item When $\alpha > \beta$,
        \begin{itemize}
            \item if $\sigma > 1$ and $\frac{1}{\sigma} > (2\theta-2)C_I-2\theta C_{AI}+2$,
            \begin{equation}
            \label{39}
                (\alpha^*, \beta^*, \gamma^*)=(\frac{1}{2C_A}, \frac{C_D}{2{C_A}^2}, 0),
            \end{equation}
            \item if $\sigma > \frac{1}{1-\gamma_m}$ and $\frac{1}{\sigma}<(1-\gamma_m)((2\theta-2)C_I-2\theta C_{AI}+2)$,
            \begin{equation}
            \label{40}
                (\alpha^*, \beta^*, \gamma^*)=(\frac{1}{2C_A}, \frac{C_D}{2(1-\gamma_m){C_A}^2}, \gamma_m),
            \end{equation}
        \end{itemize}       
    \end{itemize}
with $\theta=\frac{\theta_1}{\theta_2}$, where $\theta_1 > 0$, $\theta_2 > 0$ and $\theta_1+\theta_2 < 1$.
\end{theorem}

Theorem \ref{theorem_bayes} characterizes the BNE when the belief parameters satisfy $\theta_1 > 0$, $\theta_2 > 0$, and $\theta_1+\theta_2 < 1$. Next, we consider three edge cases in the BG-FlipIn framework, in which the insider’s preference is known with certainty: a malicious insider with $\theta_1=1$, an inadvertent insider with $\theta_1=\theta_2=0$, and a corrupt insider with $\theta_2=1$. In each case, the BNE becomes the NE. By applying similar proof techniques, we obtain the following three corollaries.

\begin{corollary}
\label{theorem_malicious}
Given $\theta_1=1$, if an equilibrium profile $(\alpha^*, \beta^*, \gamma^*)$ is an NE of BG-FlipIn with a certain malicious insider, then its closed-form expression is subject to the following conditions:
    \begin{itemize}
        \item When $\alpha \leqslant \beta$,
        \begin{itemize}
            \item if $ \sigma \leqslant 1$ and $\sigma < 2C_I$,
            \begin{equation*}
            \label{24}
                (\alpha^*, \beta^*, \gamma^*)=(\frac{C_A}{2{C_D}^2}, \frac{1}{2C_D}, 0),
            \end{equation*}
            \item if $\frac{2C_I}{1-\gamma_m} < \sigma \leqslant \frac{1}{1-\gamma_m}$ ,
            \begin{equation*}
            \label{25}
                (\alpha^*, \beta^*, \gamma^*)=(\frac{C_A (1-\gamma_m)^2}{2{C_D}^2}, \frac{1-\gamma_m}{2C_D}, \gamma_m).
            \end{equation*}
        \end{itemize}
        \item When $\alpha > \beta$,
        \begin{itemize}
            \item if $\sigma > 1$ and $\frac{1}{\sigma}>2(1-C_I)$,
            \begin{equation}
            \label{26}
                (\alpha^*, \beta^*, \gamma^*)=(\frac{1}{2C_A}, \frac{C_D}{2{C_A}^2}, 0),
            \end{equation}
            \item if $ \sigma>\frac{1}{1-\gamma_m}$ and $ \frac{1}{\sigma}< 2(1-C_I)(1-\gamma_m)$,
            \begin{equation}
            \label{27}
                (\alpha^*, \beta^*, \gamma^*)=(\frac{1}{2C_A}, \frac{C_D}{2(1-\gamma_m){C_A}^2}, \gamma_m).
            \end{equation}
        \end{itemize}       
    \end{itemize}
\end{corollary}
\begin{corollary}
\label{theorem_inadvertent}
Given $\theta_1=\theta_2=0$, if an equilibrium profile $(\alpha^*, \beta^*, \gamma^*)$ is an NE of BG-FlipIn with a certain inadvertent insider, then its closed-form expression is subject to the following conditions:
   \begin{itemize}
       \item If $\alpha \leqslant \beta$ and $ \sigma \leqslant \frac{1}{1-\gamma}$,
       \begin{equation*}
           (\alpha^*, \beta^*, \gamma^*)=(\displaystyle{\frac{C_A{(1-\gamma)^2}}{2{C_D}^2}},
        \displaystyle{\frac{1-\gamma}{2C_D}}, \gamma).
       \end{equation*}
       \item If $\alpha > \beta$ and $\sigma>\frac{1}{1-\gamma}$,
       \begin{equation*}
           (\alpha^*, \beta^*, \gamma^*)=(\displaystyle{\frac{1}{2C_A}}, \displaystyle{\frac{C_D}{2(1-\gamma){C_A}^2}}, \gamma).
       \end{equation*}
   \end{itemize}   
\end{corollary}
\begin{corollary}
\label{theorem_corrupt}
Given $\theta_2=1$, if an equilibrium profile $(\alpha^*, \beta^*, \gamma^*)$ is an NE of BG-FlipIn with a certain corrupt insider, then its closed-form expression is subject to the following conditions:
    \begin{itemize}
        \item If $\alpha \leqslant \beta$ and $\sigma \leqslant \frac{1}{1-\gamma_m}$,
        \begin{equation*}
            (\alpha^*, \beta^*, \gamma^*)=(\displaystyle{\frac{C_A (1-\gamma_m)^2}{2{C_D}^2}}, 
           \displaystyle{\frac{1-\gamma_m}{2C_D}}, \gamma_m).
        \end{equation*}
        \item If $\alpha > \beta$ and $\sigma > \frac{1}{1-\gamma_m}$,
        \begin{equation*}
            (\alpha^*, \beta^*, \gamma^*)=(\displaystyle{\frac{1}{2C_A}}, 
           \displaystyle{\frac{C_D}{2(1-\gamma_m){C_A}^2}}, \gamma_m).
        \end{equation*}
    \end{itemize}
\end{corollary}
\begin{remark}
In Corollary \ref{theorem_inadvertent}, $\gamma$ represents a known constant that indicates the percentage of resources unintentionally leaked by the inadvertent insider. Moreover, when $\theta_1 = 0$, $\theta_2 = 0$, or $\theta_1+\theta_2 = 1$, at most two insider preferences exist. These cases can be analyzed like the proofs of Theorem \ref{theorem_bayes}, and we omit their detailed discussion here.
\end{remark}

\section{Decision-making guidance for defender}

In this section, based on the expressions of the BNE and the three edge-case NEs, we analyze several phenomena in the BG-FlipIn framework. We then provide decision-making guidance for the defender under different values of $\theta_1$ and $\theta_2$. Moreover, we identify a parameter interval in which the BNE strictly outperforms all corresponding NEs, offering theoretical guidance for parameter selection in the next section.

\subsection{Decision-making with edge-case NEs}

We first investigate the three NEs presented in Corollaries \ref{theorem_malicious}, \ref{theorem_inadvertent}, and \ref{theorem_corrupt} in the previous section. Let $\mU_D^*$ represent the defender's benefit when achieving NE. In the following corollaries, we will reveal that $\mU_D^*$ is a function of $\sigma$ by substituting the closed-form NE expressions. 
\begin{corollary}
\label{U_D^*_M}
Given $\theta_1=1$, consider BG-FlipIn with a certain malicious insider. If $\frac{1}{2}<C_I<1$, then the defender's benefit $\mU_D^*$ can be expressed in three cases: First, if $\alpha \leqslant \beta$ and $\sigma \leqslant 1$, $\mU_D^* = 0$; Next, if $\alpha > \beta$ and $1 < \sigma < \frac{1}{2(1 - C_I)}$, $\mU_D^* = 1 - \frac{1}{\sigma} > 0$; Finally, if $\alpha > \beta$ and $\sigma > \frac{1}{2(1 - C_I)(1 - \gamma_m)}$, $\mU_D^* = 1 - \gamma_m - \frac{1}{\sigma} > 0$.
\end{corollary}
\begin{remark}
Note that the NE \eqref{26} exists when $C_I > \frac{1}{2}$, and the NE \eqref{27} exists when $C_I < 1$. Therefore, in Corollary \ref{U_D^*_M}, we focus on discussing the interval $\frac{1}{2} < C_I < 1$, as the physical significance of this interval is important. Other intervals can be analyzed similarly. 
\end{remark}
\begin{corollary}
\label{U_D^*_I}
Given $\theta_1=\theta_2=0$, consider BG-FlipIn with a certain inadvertent insider. The defender's benefit $\mU_D^*$ can be expressed in two cases: First, if $\alpha \leqslant \beta$ and $\sigma \leqslant \frac{1}{1 - \gamma}$, $\mU_D^* = 0$; Next, if $\alpha > \beta$ and $\sigma > \frac{1}{1 - \gamma}$, $\mU_D^* = 1 - \gamma - \frac{1}{\sigma} > 0$.
\end{corollary}
\begin{corollary}
\label{U_D^*_C}
Given $\theta_2=1$, consider BG-FlipIn with a certain corrupt insider. The defender's benefit $\mU_D^*$ can be expressed in two cases: First, if $\alpha \leqslant \beta$ and $\sigma \leqslant \frac{1}{1 - \gamma_m}$, $\mU_D^* = 0$; Next, if $\alpha > \beta$ and $\sigma > \frac{1}{1 - \gamma_m}$, $\mU_D^* = 1 - \gamma_m - \frac{1}{\sigma} > 0$.
\end{corollary}

In Corollaries \ref{U_D^*_M}, \ref{U_D^*_I} and \ref{U_D^*_C}, it is clear that when the defender moves slower than the attacker, i.e., $\alpha \leqslant \beta$, the defender’s benefit $\mU_D^*$ is 0. However, when the defender moves faster than the attacker, i.e., $\alpha > \beta$, $\mU_D^*$ becomes positive. This implies that, \textbf{regardless of the insider preference, only with the higher move rate can the defender gain benefit.} This phenomenon is consistent with other studies. For example, it has been shown that a defender can reduce compromised resources by recapturing them at a higher rate \cite{hu2015dynamic}.

Following the intuitive phenomenon discussed previously, our model can also reveal a seemingly counterintuitive phenomenon, as stated in the following corollary, whose proof is in Appendix \ref{app2}:
\begin{corollary}
\label{lem_gamma_m}
Given $\theta_1=1$, consider BG-FlipIn with a certain malicious insider. If $\frac{1}{2}<C_I<1$, then regardless of the value of $\gamma_m$, $\exists \sigma_1 < \sigma_2$, s.t. $\mU_D^*(\sigma_1)>\mU_D^*(\sigma_2)$.
\end{corollary}

Corollary \ref{lem_gamma_m} shows that when facing a malicious insider, if the cost $C_I$ lies within a certain range, \textbf{an increase in the defender’s move cost $C_D$ may paradoxically yield a greater benefit $\mU_D^*$.} Actually, this counterintuitive phenomenon can be explained by General Deterrence Theory (GDT) \cite{straub1998coping}. Although a higher $C_D$ may appear disadvantageous to the defender, it reduces the malicious insider’s expected benefit and thus discourages them from stealing resources. On the other hand, inadvertent and corrupt insiders do not gain directly from the defender’s resources, and therefore their behavior is not influenced by deterrence.

\begin{figure}[t]
\centering
\subfloat[Defender's benefit impacted by low-$C_I$ malicious insider
]{\includegraphics[width=0.45\columnwidth]{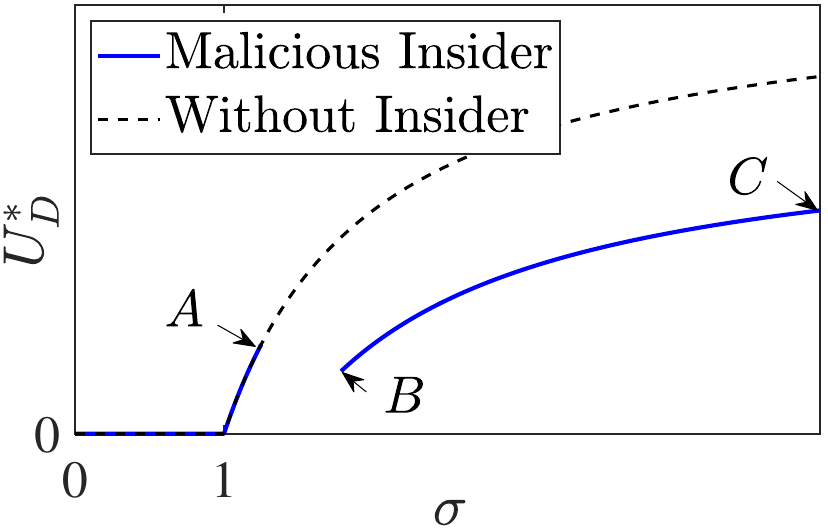}%
\label{test_M_without_insider_1}}
\vspace{0.1mm}
\subfloat[Defender's benefit impacted by high-$C_I$ malicious insider]{\includegraphics[width=0.45\columnwidth]{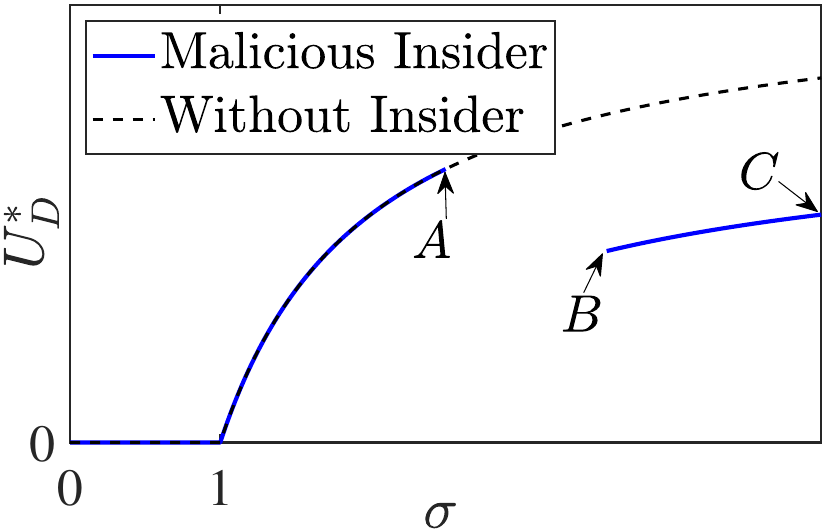}%
\label{test_M_without_insider_2}}
\caption{The defender's benefit $\mU_D^*$ impacted by malicious insider vs. the attack-defense cost ratio $\sigma$}
\label{test_M_without_insider}
\end{figure}

\begin{figure}[t]
\centering
\subfloat[Defender's benefit impacted by inadvertent insider]{\includegraphics[width=0.45\columnwidth]{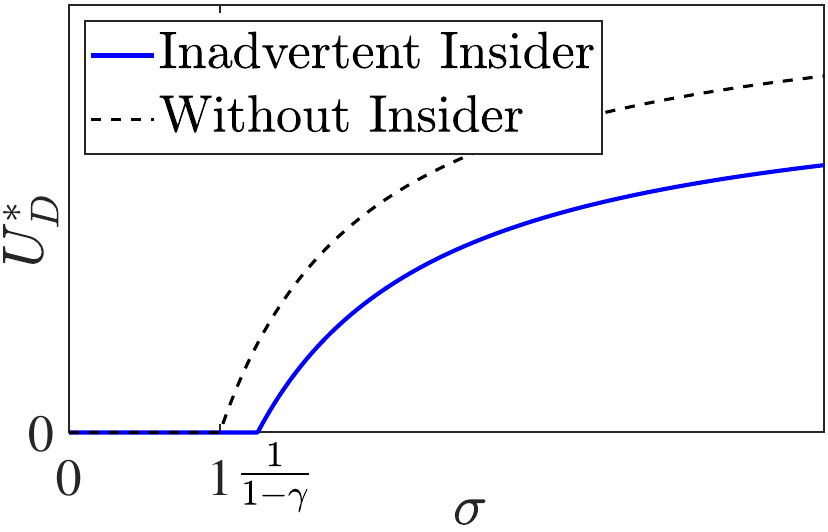}%
\label{test_I_without_insider}}
\vspace{0.1mm}
\subfloat[Defender's benefit impacted by corrupt insider]{\includegraphics[width=0.45\columnwidth]{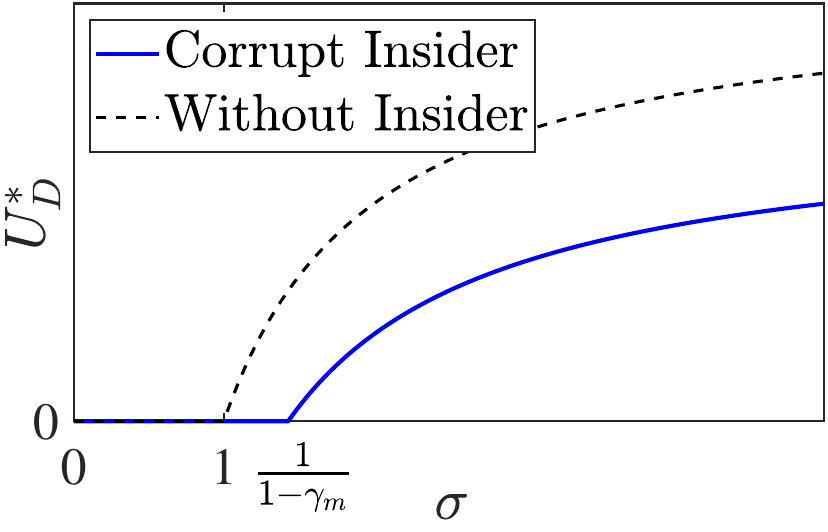}%
\label{test_C_without_insider}}
\caption{The defender's benefit $\mU_D^*$ impacted by inadvertent and corrupt insider vs. the attack-defense cost ratio $\sigma$}
\label{test_I_C_without_insider}
\vspace{-10pt}
\end{figure}

Building on the above phenomena, we provide further decision-making guidance for the defender by investigating the choice of the attack-defense cost ratio $\sigma$. By selecting an appropriate $\sigma$, the defender can maximize its benefit $\mU_D^*$ for each of the three certain insider preferences. In Figs. \ref{test_M_without_insider}, \ref{test_I_without_insider}, and \ref{test_C_without_insider}, we plot the defender's benefit defined in Corollaries \ref{U_D^*_M}, \ref{U_D^*_I} and \ref{U_D^*_C}, respectively. It is evident that the introduction of an insider reduces the defender’s benefit \(\mU_D^*\) compared to the baseline (without the insider). To minimize the harm caused by the insider, the defender is suggested to adopt distinct countermeasures for different situations. When facing a malicious insider, as illustrated in Figs. \ref{test_M_without_insider_1} and \ref{test_M_without_insider_2}, the positions of three key points are important, defined as $A:(\frac{1}{2(1-C_I)},2C_I-1)$, $B:(\frac{1}{2(1-\gamma_m)(1-C_I)},(1-\gamma_m)(2C_I-1))$, $C:(\sigma_{\text{max}},\mU_D^*(\sigma_{\text{max}}))$. There are two different scenarios based on the value of \(C_I\). In both scenarios, point $B$ is above point $A$, but the positional relationship between point $C$ and point $B$ differs. In the low-$C_I$ scenario (where the cost of the insider is small), point $C$ is above point $B$. Therefore, we suggest setting \(\sigma = \sigma_{\text{max}}\). In the high-$C_I$ scenario (where the cost of the insider is large), point $C$ is below point $B$, and we recommend setting \(\sigma = \frac{1}{2(1-C_I)}\). In Figs. \ref{test_I_without_insider} and \ref{test_C_without_insider}, as \(\sigma\) increases, \(\mU_D^*\) also increases, indicating that when facing an inadvertent insider or a corrupt insider, the defender can reduce costs and enhance efficiency to obtain greater benefit. 


\subsection{Decision-making with BNE}

To ensure the existence of all BNEs in Theorem \ref{theorem_bayes}, we focus on the interval $0<(\theta+1)C_I-\theta C_{AI}<\frac{1}{2}$ and $-1<(\theta-1)C_I-\theta C_{AI}<-\frac{1}{2}$. Intervals outside this range can be discussed similarly. Let $\tilde{\mU}_D^*$ represent the defender’s benefit when achieving BNE. Then by substituting the BNEs (\ref{37}), (\ref{38}), (\ref{39}), and (\ref{40}) into the benefit function (\ref{bayes model}), we have the following corollary. Similar to Corollaries \ref{U_D^*_M}, \ref{U_D^*_I}, and \ref{U_D^*_C}, this result also shows that the defender's benefit $\tilde{\mU}_D^*$ is positive under the same condition, i.e., when $\alpha > \beta$. 
\begin{corollary}
\label{U_D^*_B}
Consider BG-FlipIn. If $0<(\theta+1)C_I-\theta C_{AI}<\frac{1}{2}$ and $-1<(\theta-1)C_I-\theta C_{AI}<-\frac{1}{2}$, then the defender's benefit $\tilde{\mU}_D^*$ can be expressed in three cases: First, if $\alpha \leqslant \beta$ and either $\sigma < (2\theta+2)C_I - 2\theta C_{AI}$ or $\frac{(2\theta+2)C_I - 2\theta C_{AI}}{1-\gamma_m} < \sigma \leqslant \frac{1}{1-\gamma_m}$, then $\tilde{\mU}_D^* = 0$; Next, if $\alpha > \beta$, and $1< \sigma < \frac{1}{(2\theta-2)C_I - 2\theta C_{AI} + 2}$, then $\tilde{\mU}_D^* = 1 - \frac{1}{\sigma} > 0$; Finally, if $\alpha > \beta$ and $\sigma > \frac{1}{(1-\gamma_m)((2\theta-2)C_I - 2\theta C_{AI} + 2)}$, then $\tilde{\mU}_D^* = 1 - \gamma_m - \frac{1}{\sigma} > 0$.
\end{corollary}

Furthermore, similar to Corollary \ref{lem_gamma_m}, the following corollary also illustrates the phenomenon described by GDT.
\begin{corollary}
    Consider BG-FlipIn. If $0<(\theta+1)C_I-\theta C_{AI}<\frac{1}{2}$ and $-1<(\theta-1)C_I-\theta C_{AI}<-\frac{1}{2}$, then regardless of $\gamma_m$, $\exists \sigma_1 < \sigma_2$, s.t. $\tilde{\mU}_D^*(\sigma_1)>\tilde{\mU}_D^*(\sigma_2)$.
\end{corollary}

Due to space limitations, the proof is omitted here and will be included in a revised version if needed.

Next, we focus on the influence of $\theta$. Note that $\theta$ only depends on the probability that the insider is malicious and corrupt. Then we can obtain the following results, whose proof is shown in Appendix \ref{app3}:
\begin{theorem}
\label{cor5.1}
    Consider BG-FlipIn. The probability that the insider is inadvertent has no impact on all BNEs and defender’s benefit $\tilde{\mU}_D^*$.
\end{theorem}

In BG-FlipIn, the inadvertent insider is a non-strategic player whose strategies are not optimized against others’ strategies. Theorem \ref{cor5.1} reveals a phenomenon: \textbf{variations in the proportion of non-strategic players do not affect the decision-making of the rest players.} Similar invariance has also been observed, for instance, the optimal trading strategy of informed traders remains unaffected by noise traders \cite{kyle1985continuous}, while the cooperation rate of strategic players is unchanged despite increases in unconditional cooperators \cite{bo2005cooperation}.

\begin{figure}[t]
\centering
\subfloat[Defender's benefit impacted by low-$\theta$ Bayesian insider]{\includegraphics[width=0.45\columnwidth]{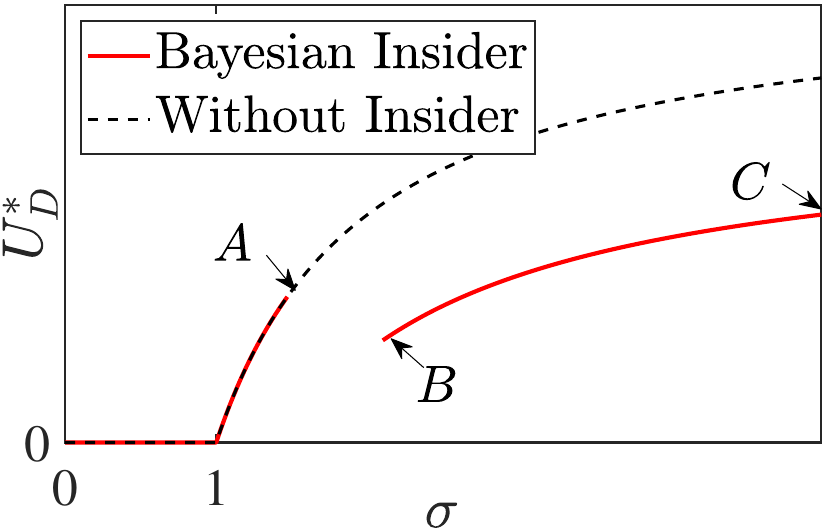}%
\label{test_B_without_insider_1}}
\vspace{0.01\textwidth}
\subfloat[Defender's benefit impacted by high-$\theta$ Bayesian insider]{\includegraphics[width=0.45\columnwidth]{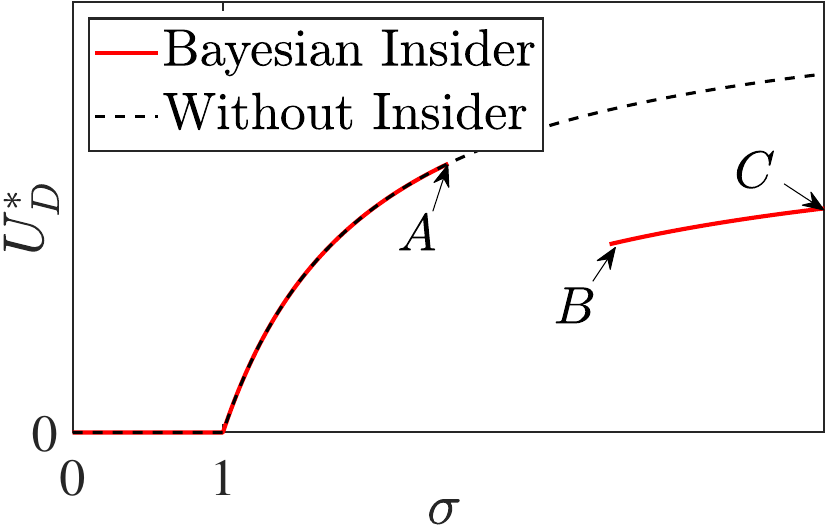}%
\label{test_B_without_insider_2}}
\caption{The defender's benefit $\tilde{\mU}_D^*$ in BG-FlipIn ($\mU_D^*$ in the FlipIt game) vs. the attack-defense cost ratio $\sigma$}
\label{test_B_without_insider}
\vspace{-10pt}
\end{figure}

Based on the above phenomena, we provide further decision-making guidance for the defender by investigating the choice of $\sigma$. By selecting an appropriate $\sigma$, the defender can maximize its benefit $\tilde{\mU}_D^*$ under uncertain insider preferences. Then with fixed $C_I$, $C_{AI}$, and $\gamma_m$, we plot the defender's benefit $\tilde{\mU}_D^*$ under two scenarios: low-$\theta$ and high-$\theta$. Moreover, we compare the differences of the defender's benefit between the FlipIt game and the Bayesian game \eqref{Γ} when achieving NE and BNE, respectively. As shown in Fig. \ref{test_B_without_insider}, the introduction of an insider of uncertain preference reduces the defender's benefit $\tilde{\mU}_D^*$ compared to the baseline (without insider). In this figure, three key points are denoted by $A\hspace{-1mm}:\hspace{-1mm}(\frac{1}{(2\theta-2)C_I-2\theta C_{AI}+2},-(2\theta-2)C_I+2\theta C_{AI}-1)$, $B:(\frac{1}{(1-\gamma_m)((2\theta-2)C_I-2\theta C_{AI}+2)},(1-\gamma_m)((2\theta-2)C_I-2\theta C_{AI}+2))$, $C:(\sigma_{\text{max}},\tilde{\mU}_D^*(\sigma_{\text{max}}))$. Point B is positioned above point A in both scenarios. However, the positional relationship between point C and point B varies. Specifically, in the low-$\theta$ scenario, point C lies above point B, while in the high-$\theta$ scenario, point C is located below point B. This variation suggests that, from the perspective of Bayesian game theory, to minimize the harm caused by the unknown preference of the insider, the defender should consider the following recommendations. Firstly, when malicious insiders predominate, the defender can achieve higher benefit by adopting $\sigma = \sigma_{\text{max}}$. On the other hand, when corrupt insiders are the majority, the defender is advised to adopt $\sigma = \frac{1}{(2\theta-2)C_I - 2\theta C_{AI} + 2}$.

\subsection{Parameter intervals ensuring BNE advantage}

In this subsection, we analyze a parameter interval for $\sigma$ to identify the conditions under which, when confronting an insider using any basic strategy, the defender can employ the Bayesian strategy to achieve greater benefit than the basic strategy. Here, the Bayesian strategy \(\alpha^*_B\) (\(\beta^*_B\), or \(\gamma^*_B\)) is referred to as the Bayesian strategy of the defender (attacker, or insider) if and only if it corresponds to the BNE within the current interval of $\sigma$ as specified in Theorem \ref{theorem_bayes}. The basic strategy $\alpha^*_k$ (or $\beta^*_k, \gamma^*_k$, where $k \in \{M, I, C\}$) is referred to as the basic strategy of the defender (attacker, or insider) if and only if it corresponds to the NE within the current interval of $\sigma$ in the BG-FlipIn with a certain malicious insider ($k=M$, Corollary \ref{theorem_malicious}), a certain inadvertent insider ($k=I$, Corollary \ref{theorem_inadvertent}), or a certain corrupt insider ($k=C$, Corollary \ref{theorem_corrupt}). This analysis explicitly maps the parameter space where the defender's Bayesian strategy outperforms all basic strategies, providing a foundation for defense strategy selection in the presence of uncertain insider threats.

In the following analysis, we consider a typical case with significant applications, where the defender and the attacker consistently adopt strategies from the same category. Specifically, both the defender and the attacker may employ basic strategy $(\alpha^*_{k_2}, \beta^*_{k_2})$, with $k_2 \in \{M, I, C\}$, or Bayesian strategy $(\alpha^*_B, \beta^*_B)$ to handle an insider using a basic strategy $\gamma^*_{k_1}$, where $k_1 \in \{M, I, C\}$. This reflects a practical situation, as both the defender and attacker are typically constrained by similar information and rational decision-making frameworks, leading them to adopt strategies from the same category.

We begin with a lemma to show that the Bayesian strategy tuples for both the defender and the attacker are essentially contained within the basic strategy tuples, without introducing additional complexity. The strategy tuples take only four forms. Moreover, we compare the defender's benefit when the defender and attacker adopt any of the four strategy tuples, under any insider strategy. Since the defender's benefit, as defined in \eqref{U_D} and \eqref{U_D_tilde}, shares the same mathematical form, the notation $U_D$ can be used here without ambiguity in the following lemma.

\begin{lemma}
\label{prop_1}
    Regardless of whether the basic strategy or Bayesian strategy is used, the defender and attacker strategy tuple only assumes four forms: For $\alpha \leq \beta$: $(\frac{C_A}{2C_D^2}, \frac{1}{2C_D})$ or $(\frac{C_A(1-\gamma)^2}{2C_D^2}, \frac{1-\gamma}{2C_D})$, for $\alpha > \beta$: $(\frac{1}{2C_A}, \frac{C_D}{2C_A^2})$ or $(\frac{1}{2C_A}, \frac{C_D}{2(1-\gamma)C_A^2})$, with $\gamma \in S_I$. Furthermore, $\forall \gamma,\gamma_0 \in S_I$, $U_D(\frac{C_A(1-\gamma)^2}{2C_D^2}, \frac{1-\gamma}{2C_D},\gamma_0)>U_D(\frac{C_A}{2C_D^2},\frac{1}{2C_D},\gamma_0)$, and $U_D(\frac{1}{2C_A},\frac{C_D}{2C_A^2},\gamma_0)>U_D(\frac{1}{2C_A},\frac{C_D}{2(1-\gamma)C_A^2},\gamma_0)$.
\end{lemma}

Lemma \ref{prop_1} follows directly from Theorem \ref{theorem_bayes} and Corollaries \ref{theorem_malicious}, \ref{theorem_inadvertent} and \ref{theorem_corrupt}. Therefore, the detailed proof is omitted.

Define the following intervals for $\sigma$:
\begin{equation*}
\label{set}
\begin{aligned}
    \mathcal{T}_M&:=\left\{\sigma\left|
\begin{array}{c}
     \frac{(2\theta+2)C_I-2\theta C_{AI}}{1-\gamma_m}<\sigma\leqslant1, \text{or}
      \\
     \frac{1}{2(1-C_I)(1-\gamma_m)}<\sigma<\frac{1}{(2\theta-2)C_I-2\theta C_{AI}+2}.
\end{array}
\right.
\right\},\\
\mathcal{T}_I&:=\left\{\sigma\left|
\begin{array}{c}
     \frac{1}{1-\gamma_m}<\sigma<\frac{1}{(2\theta-2)C_I-2\theta C_{AI}+2}.
\end{array}
\right.
\right\},\\
\mathcal{T}_C&:=\left\{\sigma\left|
\begin{array}{c}
    1<\sigma<\frac{1}{(2\theta-2)C_I-2\theta C_{AI}+2}.\\
\end{array}
\right.
\right\}.
\end{aligned}
\end{equation*}
Subsequently, based on Lemma \ref{prop_1}, we prove the following theorem, whose proof is in Appendix \ref{app4}. This theorem shows that within the intervals mentioned above, the Bayesian strategy outperforms the basic strategy. 

\begin{theorem}
\label{the_bayes_intervel}
If $\frac{1}{2}<C_I<1$, $0<(\theta+1)C_I-\theta C_{AI}<\frac{1}{2}$ and $-1<(\theta-1)C_I-\theta C_{AI}<-\frac{1}{2}$, then for all $\sigma$ within the interval $\mathcal{T}_{k_2}$, we have $U_D(\alpha^*_B,\beta^*_B,\gamma^*_{k_1})> U_D(\alpha^*_{k_2},\beta^*_{k_2},\gamma^*_{k_1})$, where $k_1,k_2\in \{M,I,C\}$. Moreover, the intersection of these intervals is non-empty, i.e., $\mathcal{T}_M \cap \mathcal{T}_I \cap \mathcal{T}_C \neq  \varnothing$.
\end{theorem}

In Theorem \ref{the_bayes_intervel}, when $\sigma$ lies within the set $\mathcal{T}_M$ ($\mathcal{T}_I$ or $\mathcal{T}_C$), we observe that, for any preference of insider, the defender's benefit from employing Bayesian strategy $\alpha^*_B$ is always greater than that obtained with basic strategy $\alpha^*_M$ ($\alpha^*_I$ or $\alpha^*_C$). Furthermore, the fact that the intersection of these three intervals is non-empty indicates that there exists a range of $\sigma$ where the defender, facing any preference of the insider, can achieve greater benefit by using the Bayesian strategy over all basic strategies.
\begin{figure}[t]
    \centering
\subfloat[Malicious strategy vs. Bayesian strategy ]{\includegraphics[width=1\linewidth]{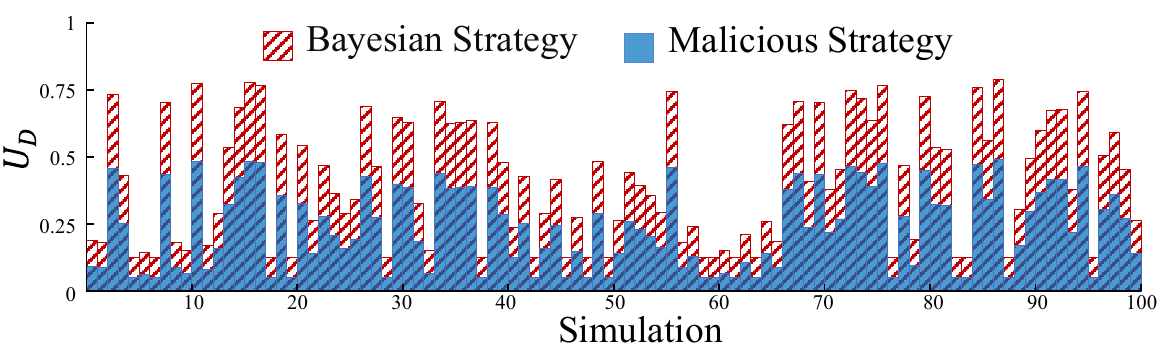}%
\label{1_1}}
\newline
\subfloat[Inadvertent strategy vs. Bayesian strategy]{\includegraphics[width=1\linewidth]{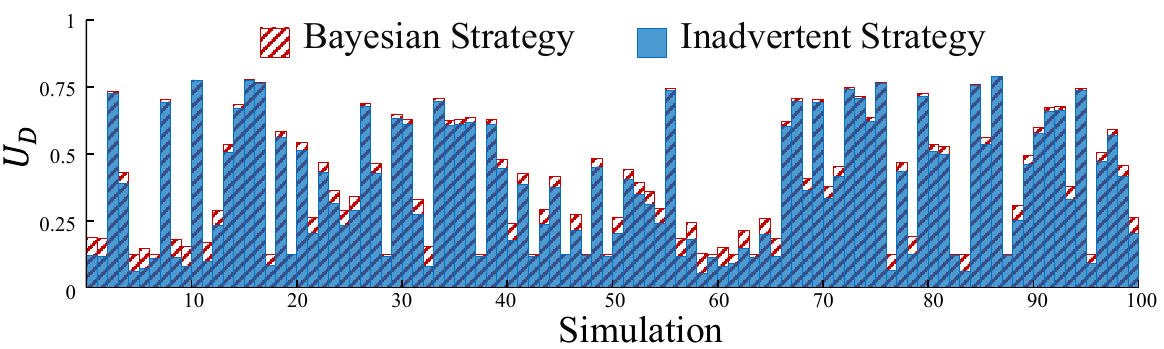}%
\label{1_2}}
\newline
\subfloat[Corrupt strategy vs. Bayesian strategy]{\includegraphics[width=1\linewidth]{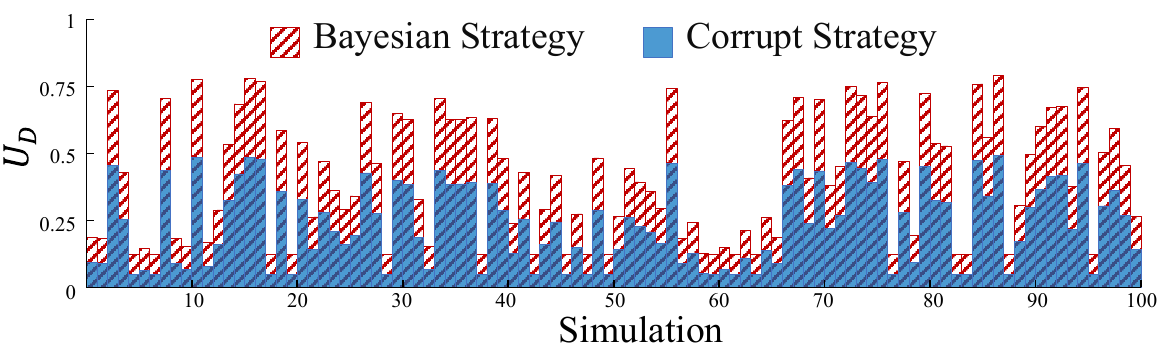}%
\label{1_3}}
\caption{The comparison of malicious, inadvertent, corrupt and Bayesian strategy under unknown insider preference, $C_D=0.2,C_I=0.51,C_{AI}=1.02,\gamma_m=0.75,\theta_1=\theta_2=0.1$}
    \label{test_C_I_W_vs_Bayes}
\vspace{-5pt}
\end{figure}

\section{Application}

In this section, we present two applications to illustrate the significance of the BG-FlipIn when dealing with insider threats. The first application is a small-scale simulation with man-made data, aiming to examine the model's effectiveness when the insider preference is unknown. The second application is a cloud-based validation, which focuses on scenarios where the insider's preferences change rapidly in practice. To simplify parameterization, we set $C_A=1$.

\subsection{Simulation with unknown insider preferences}

In this subsection, we compare the defender's benefit when the defender and attacker use a Bayesian strategy versus a basic strategy, under the condition that the preference of the insider remains unknown. 

\textbf{1) Setup:} The simulations are implemented in MATLAB R2018b on a PC with the Intel Core i5-10210U CPU processors (2.11GHz) and 8 GB of physical memory. With fixed parameters $C_D$, $C_I$, $\gamma_m$, $\theta_1$, and $\theta_2$, we conduct 100 simulations. In each simulation, the insider's preference $t_{I}$ is randomly generated according to the insider distribution. Based on the insider's preference, it adopts the corresponding basic strategy. If the insider is inadvertent, the proportion of the resource impacted by the insider is randomly drawn from a uniform distribution over $(0, \gamma_m)$.

\textbf{2) Method:} Faced with these 100 simulations where the insider's preference is randomly determined, the defender and the attacker, lacking knowledge of the insider's preference, have to adopt a single strategy throughout all simulations. That is, across all simulations, both the defender and the attacker consistently employ either the Bayesian strategy in Theorem \ref{theorem_bayes} or one of the three basic strategies in Corollaries \ref{theorem_malicious}, \ref{theorem_inadvertent}, and \ref{theorem_corrupt}. In addition, we assume that if the preference of insider is inadvertent, the defender can identify the percentage of the resource impacted by the insider.

\begin{figure}[t]
    \centering
\subfloat[Malicious strategy vs. Bayesian strategy]{\includegraphics[width=1\linewidth]{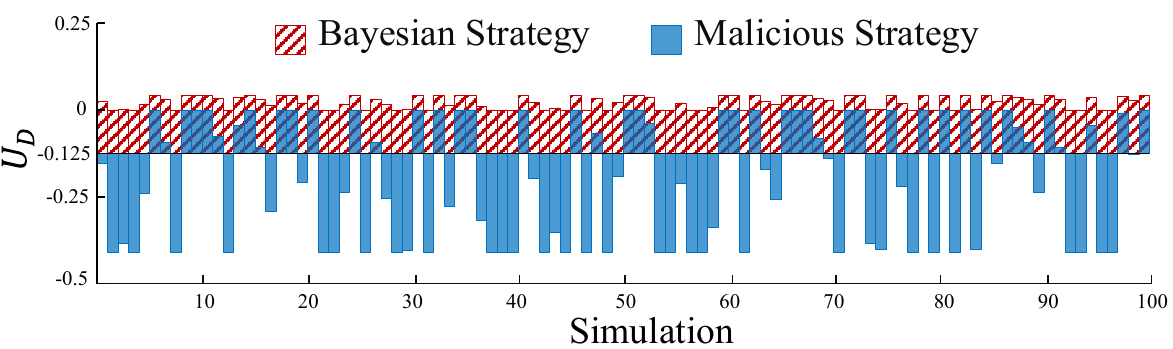}%
\label{1_1}}
\newline
\subfloat[Inadvertent strategy vs. Bayesian strategy]{\includegraphics[width=1\linewidth]{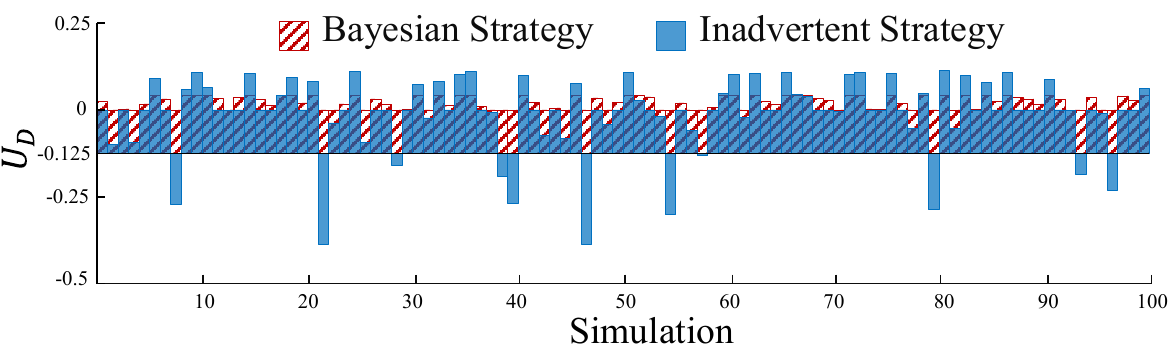}%
\label{test_I_vs_B_2}}
\newline
\subfloat[Corrupt strategy vs. Bayesian strategy]{\includegraphics[width=1\linewidth]{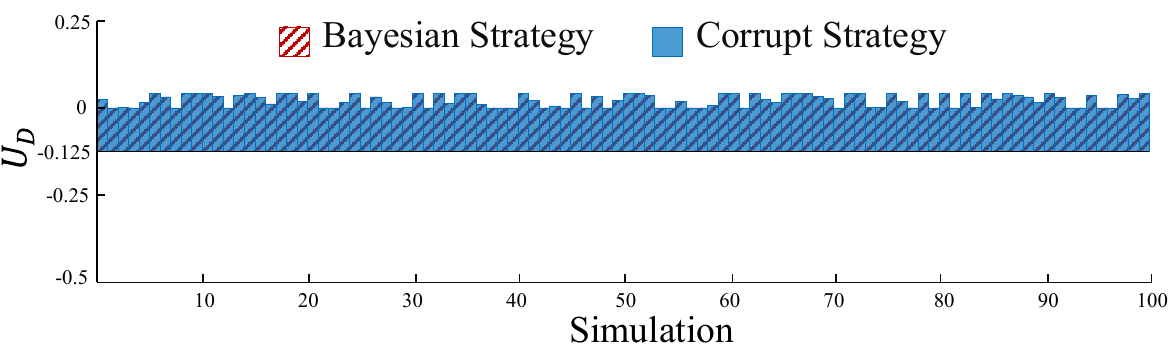}%
\label{1_3}}
\caption{The comparison of malicious, inadvertent, corrupt and Bayesian strategy under unknown insider preference, $C_D=1.1,C_I=0.99,C_{AI}=1.98,\gamma_m=0.9,\theta_1=\theta_2=0.33$}
    \label{test_C_I_W_vs_Bayes_2}
\vspace{-5pt}
\end{figure}

\textbf{3) Result:} In Fig. \ref{test_C_I_W_vs_Bayes}, we select parameters from the intersection of $\mathcal{T}_M$, $\mathcal{T}_I$, and $\mathcal{T}_C$ as proposed in Theorem \ref{the_bayes_intervel}. Under these conditions, for any preference of the insider, the Bayesian strategy consistently provides greater benefit for the defender than the other basic strategies. As shown in the figure, the red striped bars represent the defender's benefit gained using the Bayesian strategy, while the blue translucent bars represent the defender's benefit from the basic strategies. Across all simulations, the red striped bars are consistently higher than the blue translucent bars, indicating that within a specific parameter range, the Bayesian strategy outperforms the basic strategies against an unknown insider preference.

In Fig. \ref{test_C_I_W_vs_Bayes_2}, with the remaining experimental setup unchanged, we altered the parameter settings so that they fall outside the intersection of $\mathcal{T}_M$, $\mathcal{T}_I$, and $\mathcal{T}_C$ defined in Theorem \ref{the_bayes_intervel}. As shown in Fig. \ref{test_C_I_W_vs_Bayes_2}, under these parameters, the Bayesian strategy outperforms the malicious and corrupt strategies but falls short compared to the inadvertent strategy. However, by summing up the results of the 100 simulations in Fig. \ref{test_I_vs_B_2}, we find that the total benefit obtained using the Bayesian strategy is 2.1577
, while that using the inadvertent strategy is -0.7835. This further illustrates an advantage of the Bayesian strategy: although it may underperform other strategies in some simulations, it achieves better expected performance.

\subsection{Evaluation in remote state estimation}

In this subsection, we evaluate the rapid response capability of Bayesian strategies against rapidly changing insider preferences in the context of cloud-enabled remote state estimation (RSE). RSE serves as an indispensable functional module in CPS. Recent studies have shown that adversarial agents can tamper with data packets transmitted over unreliable channels (e.g., cloud infrastructures exposed to APT attacks) in RSE, which may result in significant degradation of estimation performance \cite{guo2016optimal,zhou2023eavesdropping}. 

\textbf{1) Setup:}
The validation is deployed on Amazon Web Services using Elastic Compute Cloud instances. Specifically, three t3.medium instances, each equipped with 2 vCPUs (Intel Xeon Platinum, 2.5 GHz) and 4 GB of RAM running Ubuntu 20.04 LTS are employed. 

\begin{figure*}[t]
    \centering
    \includegraphics[width=1\linewidth]{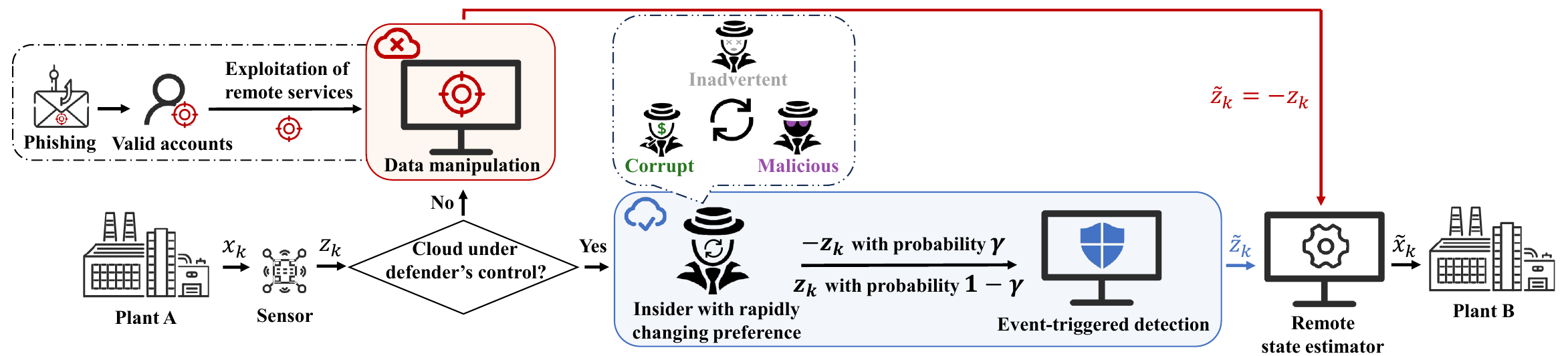}
    \caption{Architecture for APT against remote state estimation}
    \label{cloud}
\vspace{-5pt}
\end{figure*}
In the scenario when the probabilities of each insider preference are equal, we conduct four experiments using the remote state estimation model and analyze the efficacy of the Bayesian strategy compared to the basic strategy when facing with the rapidly changing preferences of insiders. The system architecture (Fig. \ref{cloud}) features a linear time-invariant process:
\begin{equation*}
\begin{split}
x_{k+1} &= A x_{k} + \omega_k, \\
y_k &= C x_k + v_k,
\end{split}
\end{equation*}
where \(k \in \mathbb{N}\) is the time index, \(x_k \in \mathbb{R}^n\) is the system state, \(y_k \in \mathbb{R}^m\) is the sensor measurement, and \(\omega_k \in \mathbb{R}^n\), \(v_k \in \mathbb{R}^m\) are zero-mean i.i.d. Gaussian noises.  The initial state \(x_0\) is Gaussian and independent of \(\omega_k\), \(v_k\).  The pair \((A, C)\) is observable, and \(\text{rank}(C) = m\).

\begin{table}[t]
    \centering
    \caption{The attacker's APT techniques}
    \begin{tabular}{c|c}
        \toprule
        Tactics & Techniques \\
        \midrule
        Initial Access (TA0001) &     Phishing (T1566)   \\
        Privilege Escalation (TA0004)&      Valid Accounts (T1078)  \\
        Lateral Movement (TA0008)   &  Remote Services (T1210)   \\
        Exfiltration (TA0010)   &   Data Manipulation (T1565)  \\
        \bottomrule
    \end{tabular}
    \label{tab:attack_mapping}
\vspace{-5pt}
\end{table}

In Fig. \ref{cloud}, the innovation sequence \(z_k = y_k - C \hat{x}_{k|k-1}\) (where \(\hat{x}_{k|k-1}\) is the prior state estimate from the Kalman filter) is central to the threat model. The state estimate of the remote estimator follows $\tilde{x}_{k} =A \tilde{x}_{k-1}+ K_k \tilde{z}_k$, where $K_k$ is the Kalman gain. The attacker periodically employs APT techniques to compromise and gain control of the cloud infrastructure. Table \ref{tab:attack_mapping} maps these techniques to the corresponding MITRE ATT\&CK framework \cite{strom2018mitre}. Conversely, the defender executes countermeasures at scheduled intervals to patch vulnerabilities and regain control. When the cloud is under attacker control, the attacker performs the optimal linear stealthy attack \cite{guo2016optimal}, replacing the output \(\tilde{z}_k = -z_k\). When the cloud is under defender control, the defender performs event-triggered detection to ensure input/output consistency \cite{eslami2023detection}, but cannot validate the correctness of $\tilde{z}_k$. During these periods, insiders alter the input \(z_k\) with probability \(\gamma\).
\begin{itemize}
    \item Malicious insider: Deliberately sets the input to \(-z_k\).
    \item Inadvertent insider: Accidentally sets the input to \(-z_k\).
    \item Corrupt insider: Opens a backdoor allowing the attacker to set the input to \( -z_k\).
\end{itemize}  
The defender’s consistency check fails to detect these alterations since \(\tilde{z}_k=-z_k\) is accepted as a valid output. 

\begin{figure*}[t]
    \centering
\subfloat[The benefit of the defender]{\includegraphics[width=0.245\linewidth]{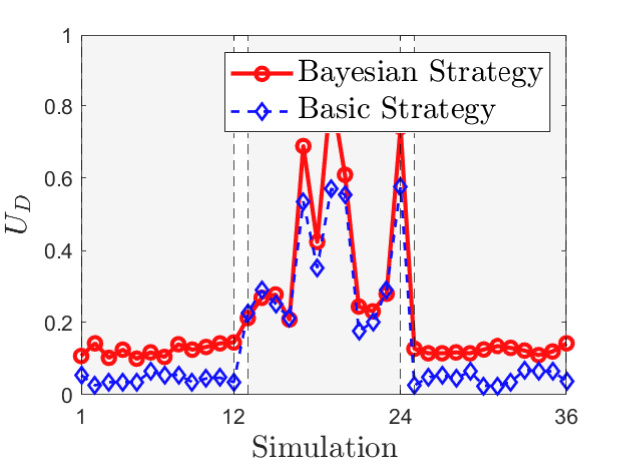}%
\label{exp_1_a}}
\subfloat[The sum of the benefit]{\includegraphics[width=0.245\linewidth]{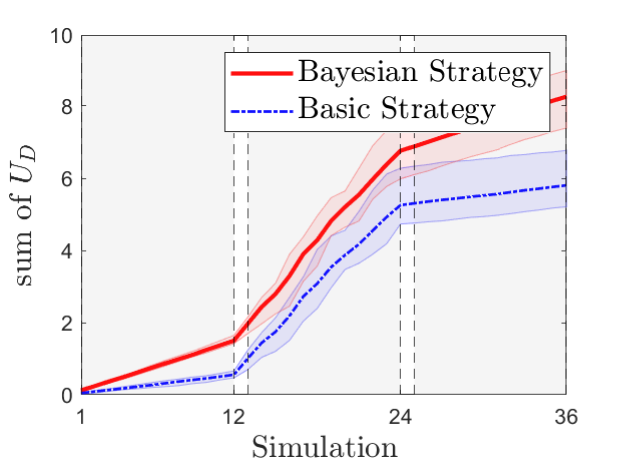}%
\label{exp_1_b}}
\subfloat[RMSE of remote state estimation]{\includegraphics[width=0.245\linewidth]{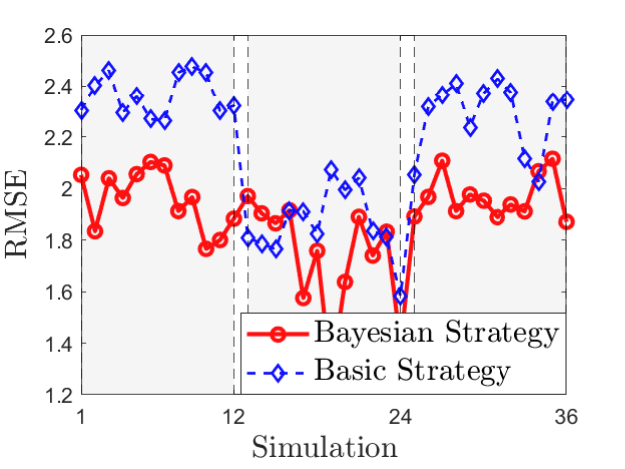}%
\label{exp_1_c}}
\subfloat[The sum of RMSE]{\includegraphics[width=0.245\linewidth]{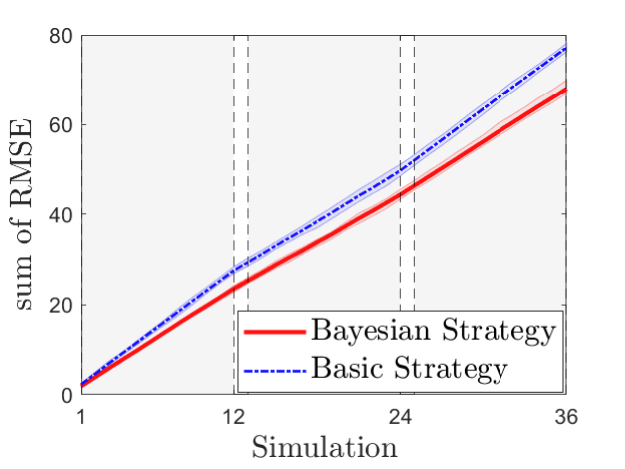}%
\label{exp_1_d}}
\caption{Numerical results of the first experiment with a $100\%$ alignment ratio}
    \label{exp_1}
\end{figure*}
\begin{figure*}[t]
    \centering
\subfloat[The benefit of the defender]{\includegraphics[width=0.245\linewidth]{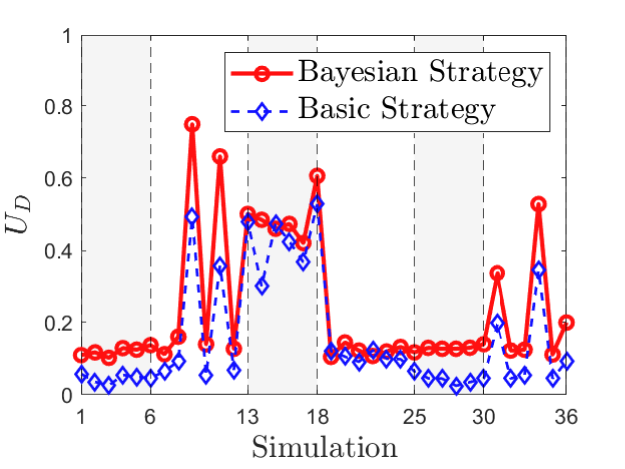}%
\label{exp_2_a}}
\subfloat[The sum of the benefit]{\includegraphics[width=0.245\linewidth]{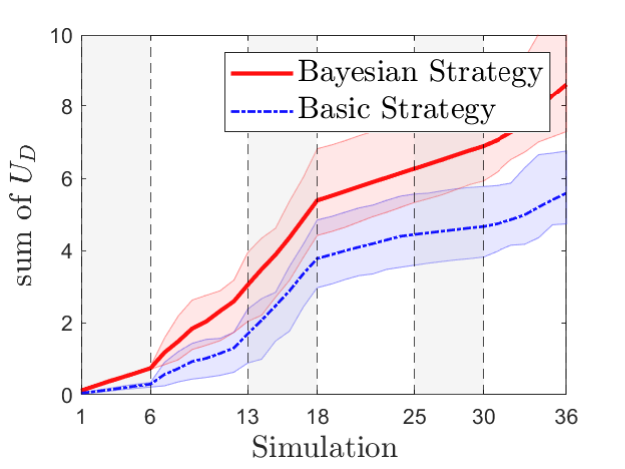}%
\label{exp_2_b}}
\subfloat[RMSE of remote state estimation]{\includegraphics[width=0.245\linewidth]{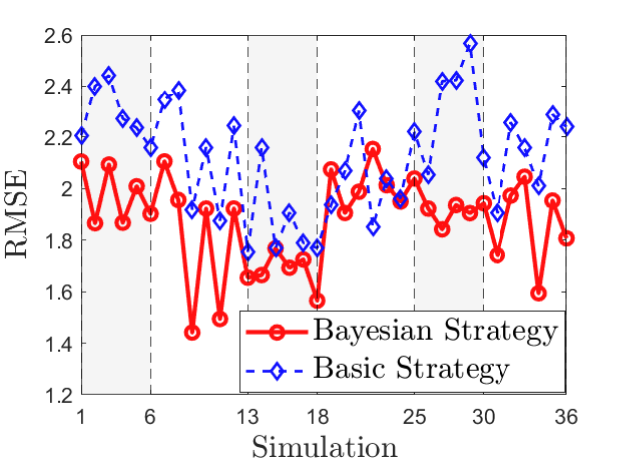}%
\label{exp_2_c}}
\subfloat[The sum of RMSE]{\includegraphics[width=0.245\linewidth]{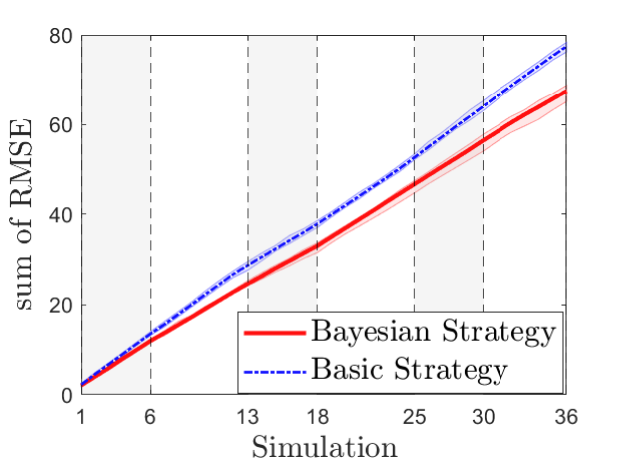}%
\label{exp_2_d}}
\caption{Numerical results of the second experiment with a $50\%$ alignment ratio}
    \label{exp_2}
\end{figure*}
\begin{figure*}[t]
    \centering
\subfloat[The benefit of the defender]{\includegraphics[width=0.245\linewidth]{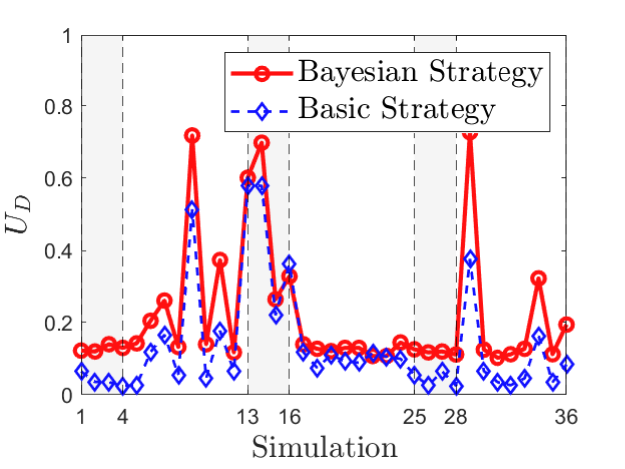}%
\label{exp_3_a}}
\subfloat[The sum of the benefit]{\includegraphics[width=0.245\linewidth]{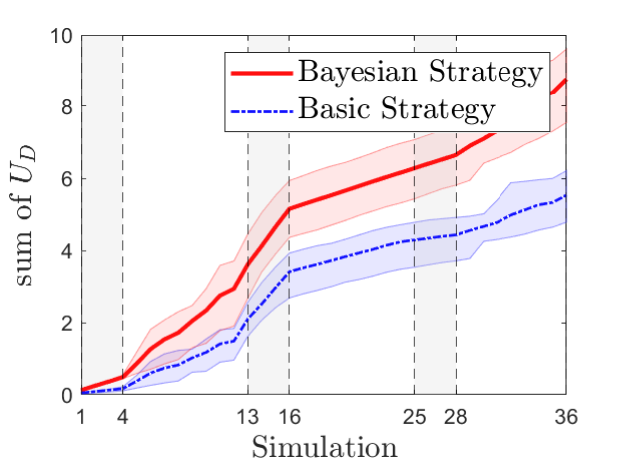}%
\label{exp_3_b}}
\subfloat[RMSE of remote state estimation]{\includegraphics[width=0.245\linewidth]{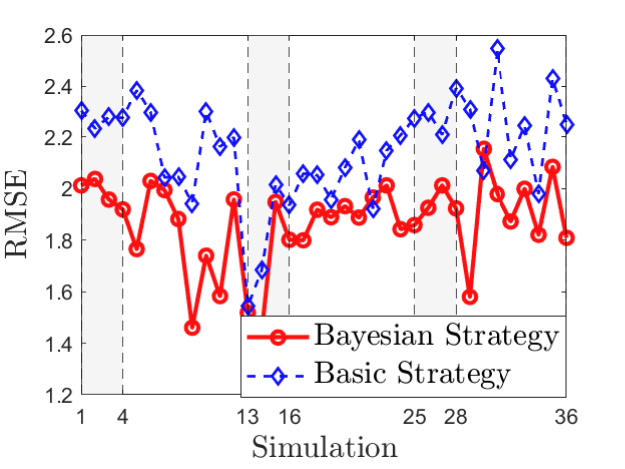}%
\label{exp_3_c}}
\subfloat[The sum of RMSE]{\includegraphics[width=0.245\linewidth]{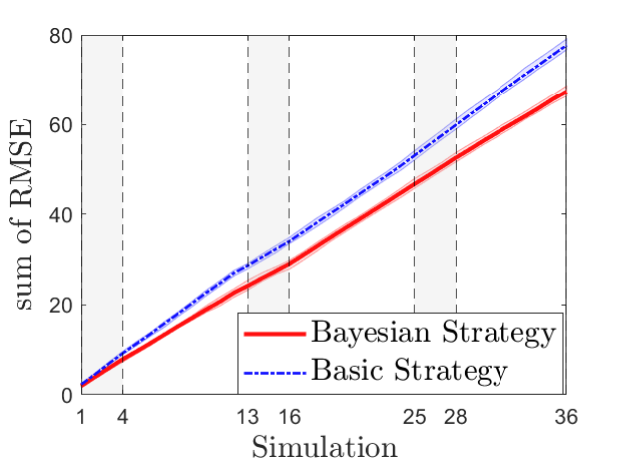}%
\label{exp_3_d}}
\caption{Numerical results of the third experiment with a $33\%$ alignment ratio}
    \label{exp_3}
\end{figure*}
\begin{figure*}[t]
    \centering
\subfloat[The benefit of the defender]{\includegraphics[width=0.245\linewidth]{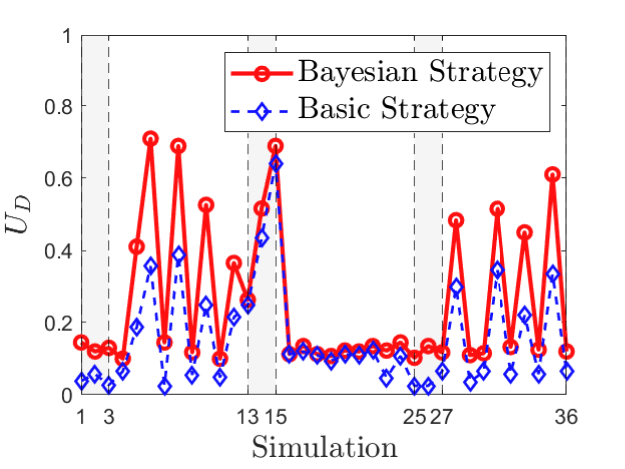}%
\label{exp_4_a}}
\subfloat[The sum of the benefit]{\includegraphics[width=0.245\linewidth]{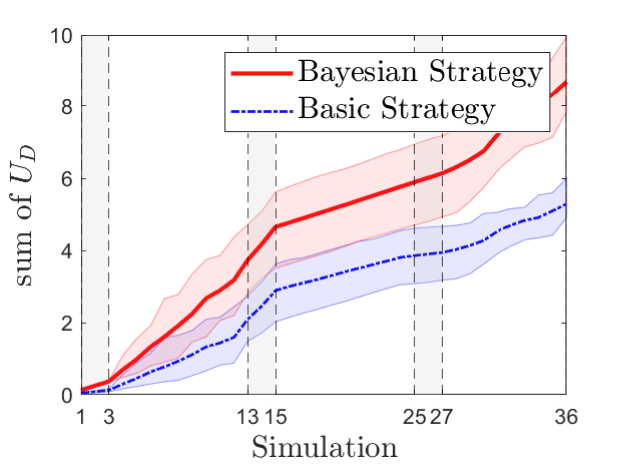}%
\label{exp_4_b}}
\subfloat[RMSE of remote state estimation]{\includegraphics[width=0.245\linewidth]{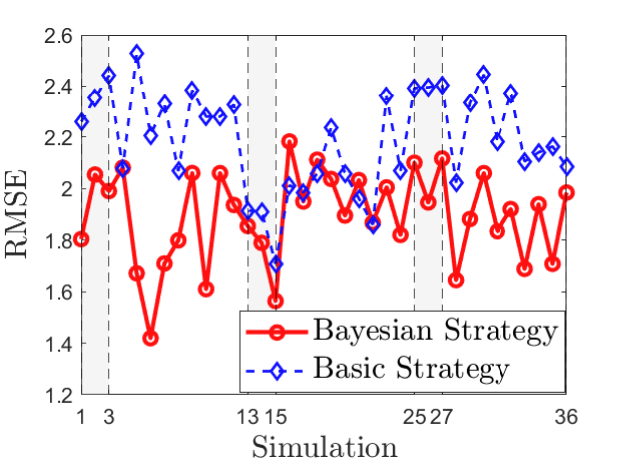}%
\label{exp_4_c}}
\subfloat[The sum of RMSE]{\includegraphics[width=0.245\linewidth]{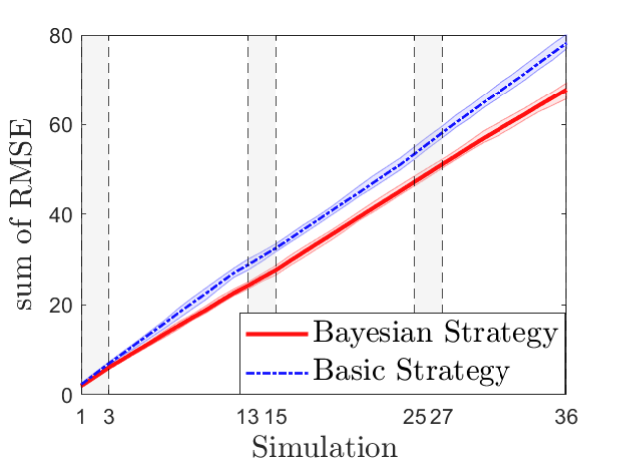}%
\label{exp_4_d}}
\caption{Numerical results of the fourth experiment with a $25\%$ alignment ratio}
    \label{exp_4}
\vspace{-10pt}
\end{figure*}

Following the progress from the first experiment to the last experiment, the variation in insider preferences becomes increasingly disordered. Specifically, each experiment consists of 36 simulations. In the $i$-th experiment, where $i=1,2,3,4$, for simulations 1 to $\frac{12}{i}$, the insider is malicious, for simulations 13 to $\frac{12}{i}+12$, the insider is inadvertent, for simulations 1 to $\frac{12}{i}+24$, the insider is corrupt. For the remaining $36 \left( \frac{i-1}{i} \right)$ simulations, the insider preferences are randomly assigned with the following constraints: in simulation $k$, where $k \in \left[ \frac{12}{i}+1, \ldots, 12 \right]$, the insider is inadvertent (or corrupt); in simulation $k+12$, the insider is corrupt (or malicious); and in simulation $k+24$, the insider is malicious (or inadvertent). If the insider is inadvertent, the proportion of the resource impacted by the insider is randomly drawn from a uniform distribution over $(0, \gamma_m)$.

\textbf{2) Method:}
In each experiment, the Bayesian strategy for the defender and the attacker is to adopt the tuple $(\alpha^*_B,\beta^*_B)$ from Theorem \ref{theorem_bayes}, and the basic strategy is to adopt $(\alpha^*_M,\beta^*_M)$ from Corollary \ref{theorem_malicious} in the first 12 simulations, $(\alpha^*_I,\beta^*_I)$ from Corollary \ref{theorem_inadvertent} in the next 12 simulations, and $(\alpha^*_C,\beta^*_C)$ from Corollary \ref{theorem_corrupt} in the final 12 simulations. Then in the $i$-th experiment, where $i=1,2,3,4$, we ensure that the alignment ratio between insider preferences and basic strategy type is $\frac{1}{i}$.

In each simulation, we consider a stable process with parameters $A = 0.8, C = 1.2, Q = 1, R = 1$, and the simulation horizon is set to $T = 100$ with a fixed sampling interval $\Delta t = 0.1$. Let $T_D$ represent the total time during which the cloud is under the defender's control in the simulation, and let $N=\frac{T}{\Delta t}$. Calculate the benefit $U_D$ of the defender and the root mean square error (RMSE) of remote state estimation, where
$
U_D=\frac{T_D}{T}(1-\gamma)-C_D\alpha,
$
and
$
\text{RMSE}=\sqrt{\frac{1}{N}\sum_{k=1}^{N}(x_k-\tilde{x}_k)(x_k-\tilde{x}_k)^{T}}.
$
Additionally, we calculate the cumulative sum of $U_D$ and the cumulative sum of the RMSE across simulation indices. 

\textbf{3) Result:}
All the results in four experiments have been plotted
with respect to the simulation index, as shown
in Figs. \ref{exp_1}, \ref{exp_2}, \ref{exp_3}, and \ref{exp_4}. In each figure, sub-figure (a) illustrates the individual $U_D$ values for each simulation, while sub-figure (b) displays the cumulative sum of $U_D$, highlighting the overall trend in defender benefit accumulation. Similarly, sub-figures (c) and (d) show the RMSE for each simulation and its cumulative progression, respectively. Moreover, we record the total $U_D$ and total RMSE aggregated over all 36 simulations in each experiment, as summarized in Tab. \ref{tab2}.

In Figs. \ref{exp_1}, \ref{exp_2}, \ref{exp_3}, and \ref{exp_4}, we deliberately choose parameters outside the intersection of $\mathcal{T}_M$, $\mathcal{T}_I$, and $\mathcal{T}_C$ as specified in Theorem \ref{the_bayes_intervel}: $C_D=0.2, C_I=0.51,C_{AI}=1.01, \gamma_m=0.75,\theta_1=\theta_2=0.33$. This setting may cause the Bayesian strategy to underperform the basic strategy at certain points. However, our results show that as the insider's preference change more rapidly, the Bayesian strategy performs increasingly better. From the cumulative plots, it becomes evident that the Bayesian strategy consistently maintains a higher cumulative $U_D$ and lower RMSE compared to the basic strategies. This advantage becomes more evident in later experiments (e.g., Experiments 3 and 4) where the insider's preferences change more rapidly and unpredictably.

Tab. \ref{tab2} quantitatively confirms this trend. Across all four experiments, the Bayesian strategy yields significantly greater total $U_D$ compared with the basic strategy (e.g., 8.6958 vs. 5.2844 in Experiment 4) and lower total RMSE compared with the basic strategy (e.g., 67.7650 vs. 78.1761 in Experiment 4). Notably, the performance gap between the Bayesian and basic strategies widens as the volatility of insider behavior increases, which substantiates the BG-FlipIn's capacity to cope with rapid shifts without recognition of the insider preference.

These findings collectively underscore the necessity of the Bayesian framework in practical APT defense when insider preferences are uncertain and even time-varying.

\begin{table}[h]
    \centering
    \caption{Total $\mU_D$ and RMSE under basic and Bayesian strategies for different alignment ratios}
    \renewcommand{\arraystretch}{1.1}
    \label{tab2}
    \begin{tabular}{|c|c|c|c|c|c|}
        \hline
        \multirow{2}{*}{Metric} & \multirow{2}{*}{Strategy} & \multicolumn{4}{c|}{Alignment ratio} \\
        \cline{3-6}
        & & 100\% & 50\% & 33\% & 25\% \\
        \hline
        \multirow{3}{*}{$U_D$} 
        & Basic & 5.8028 & 5.5865 & 5.5317 & 5.2844 \\
        \cline{2-6}
        & Bayesian & 8.2609 & 8.6249 & 8.6689 & 8.6958 \\
        \cline{2-6}
        & Difference & \textbf{+2.4581} & \textbf{+3.0384} & \textbf{+3.1372} & \textbf{+3.4114} \\
        \hline
        \multirow{3}{*}{RMSE} 
        & Basic  & 77.1941 & 77.4638 & 77.7119 & 78.1761 \\
        \cline{2-6}
        & Bayesian  & 67.9569 & 67.4668 & 67.3191 &  67.7650 \\
        \cline{2-6}
         & Difference & \textbf{-9.2372} & \textbf{-9.9970} & \textbf{-10.3928} & \textbf{-10.4111} \\
        \hline
    \end{tabular}
   
\end{table}

% \vspace{-10pt} 
\section{Conclusion}

In this paper, we proposed BG-FlipIn: a Bayesian game framework for FlipIt-insider models that investigates malicious, inadvertent, and corrupt insiders. We then derived the BNE and analyzed three edge cases with certain insider preferences to obtain the corresponding NE. Based on BNE and NEs, we discovered several phenomena related to the defender’s move rate and cost, and the insider's preferences. We then provided decision-making guidance for the defender under both certain and uncertain insider preferences. Moreover, we identified a parameter interval in which the BNE offered an advantage. Finally, two applications were presented to illustrate the performance and significance of BG-FlipIn in dealing with insider threats. 

% Our results highlighted the value of probabilistic modeling in addressing uncertainty in insider preferences, providing an approach for the defender to make informed decisions while reducing the cost of preference detection and avoiding frequent strategy adjustments.

\appendices
\section{The proof of the Theorem \ref{theorem_bayes} \label{app1}}

We first presume that the Bayesian game for the FlipIt-insider model (\ref{Γ}) possesses a BNE denoted as $(\alpha^*, \beta^*, \gamma^*)$.

When $\alpha \leqslant \beta$, the benefit functions (\ref{bayes model}) of the Bayesian game (\ref{Γ}) can be reformulated as follows:
\begin{equation*}\left\{
\begin{aligned}
&\tilde{\mU}_D = \alpha F, \\
&\tilde{\mU}_A = 1 - \frac{\alpha}{2\beta} - C_A\beta - \theta_2 C_{AI}\gamma, \\
&\tilde{\mU}_I = \gamma H,
\end{aligned}
\right.
\end{equation*}
where we define
$$F= \frac{1-\gamma}{2\beta} - C_D,\quad H= \theta_1(x-C_I) + \theta_2(C_{AI}-C_I).$$ 

Since $F$ is independent of $\alpha$, the defender’s benefit function $\tilde{\mU}_D$ is linear in $\alpha$. Hence, when $F \neq 0$, the maximum benefit is attained at the boundary, i.e., $\alpha^* = 0$ if $F < 0$, and $\alpha^* = \alpha_m$ if $F > 0$. Similarly, because $H$ is independent of $\gamma$, the insider’s benefit function $\tilde{\mU}_I$ is linear in $\gamma$. Thus, when $H \neq 0$, the maximum benefit occurs at $\gamma^* = 0$ if $H < 0$, and $\gamma^* = \gamma_m$ if $H > 0$. 

In contrast, the attacker’s benefit function $\tilde{\mU}_A$ is not linear in $\beta$, and therefore $\beta^*$ cannot be determined in the same way as $\alpha^*$ and $\gamma^*$. Instead, we observe that when $\alpha = 0$, the partial derivative of $\tilde{\mU}_A$ with respect to $\beta$ reduces to a negative constant $-C_A$. When $\alpha$ is treated as a nonzero constant, the derivative is $$\frac{\mathrm{d}\tilde{\mU}_A}{\mathrm{d}\beta}=\frac{\alpha}{2\beta^2}-C_A,$$
which is strictly decreasing in $\beta$ and admits a unique zero point at 
$$\beta_0=\sqrt{\frac{\alpha}{2C_A}}.$$ 
Therefore, $\beta^* = 0$ if $\alpha^* = 0$, and $\beta^* = \beta_0$ if $\alpha^* \neq 0$.

Subsequently, we focus on the following five cases:

1) If $F < 0$, then $\alpha^* = 0$, and $\beta^* = 0$, with $\gamma^* \to 1$, but since $\gamma^*$ is not greater than $\gamma_m$, there is no valid equilibrium for this case.

2) If $F > 0$, then $\alpha^* = \alpha_m$, and $\beta^* = \beta_0 = \sqrt{\frac{\alpha_m}{2C_A}}$, but this leads to a contradiction, as $\alpha^*$ cannot be greater than $\beta^*$ in this case. Therefore, this case does not yield a valid equilibrium either.

3) If $F = 0$ and $H<0$, then $\gamma^*=0$. From $F=0$, we obtain $$(\alpha^*, \beta^*, \gamma^*) = (\alpha, \sqrt{\frac{\alpha}{2C_A}}, 1-2C_D\sqrt{\frac{\alpha}{2C_A}}), \forall \alpha \in S_D.$$ This simplifies to $$(\alpha^*, \beta^*, \gamma^*) = (\frac{C_A}{2C_D^2}, \frac{1}{2C_D}, 0).$$ If this triplet satisfies $H(\alpha^*, \beta^*, \gamma^*) < 0$ and $\alpha^* \leqslant \beta^*$, it constitutes a BNE. 

4) If $F = 0$ and $H>0$, then $\gamma^* = \gamma_m$. Similarly, we have $$(\alpha^*, \beta^*, \gamma^*) = (\frac{C_A(1-\gamma_m)^2}{2C_D^2}, \frac{1-\gamma_m}{2C_D}, \gamma_m).$$ If this triplet fulfills $H(\alpha^*, \beta^*, \gamma^*) > 0$ and $\alpha^* \leqslant \beta^*$, it represents a BNE. 

5) If $F = 0$ and $H=0$, the solution obtained by $F=0$ and $H=0$ has measure zero, so this case is not considered.

Next, when $\alpha > \beta$, the benefit functions can be written as
\begin{equation*}\left\{
\begin{aligned}
&\tilde{\mU}_D = (1-\gamma)(1 - \frac{\beta}{2\alpha}) - C_D\alpha, \\
&\tilde{\mU}_A = \beta K - \theta_2 C_{AI} \gamma, \\
&\tilde{\mU}_I = \gamma H,
\end{aligned}\right.
\end{equation*}
where $$K = \frac{1}{2\alpha} - C_A, \quad H=\theta_1(x-C_I) + \theta_2(C_{AI}-C_I).$$
Since $\tilde{\mU}_A$ is linear in $\beta$ for fixed $\alpha$ and $\gamma$, if $K \neq 0$, the attacker’s maximum benefit is attained at the boundary: $\beta^* = 0$ when $K < 0$, and $\beta^* = \beta_m$ when $K > 0$. Similarly, if $H < 0$, the insider will choose $\gamma^* = 0$, and if $H > 0$, $\gamma^* = \gamma_m$. 

The defender strategy $\alpha^*$ depends on the attacker’s choice of $\beta$. If $\beta = 0$, then clearly $\alpha^* = 0$. When $\beta$ is treated as a positive constant, the partial derivative of $\tilde{\mU}_D$ with respect to $\alpha$ is
$$ \frac{\mathrm{d}\tilde{\mU}_D}{\mathrm{d}\alpha}=(1-\gamma)\frac{\beta}{2\alpha^2}-C_D,$$
which is strictly decreasing in $\alpha$. Then $\alpha^*$ is given by the zero point of the derivative, i.e., $$\alpha^*=\alpha_0 = \sqrt{\frac{(1-\gamma)\beta}{2C_D}}.$$

Subsequently, we focus on the following two cases:

1) If $K < 0$, then $\beta^* = 0$. From $\alpha^* = \alpha_0 = \sqrt{\frac{(1-\gamma)\beta}{2C_D}}$, we have $\alpha^* = 0$, which contradicts the condition $\alpha > \beta$. Therefore, this case does not yield a valid equilibrium.

2) If $K > 0$, then $\beta^* = \beta_m$. From $\alpha^* = \alpha_0 = \sqrt{\frac{(1-\gamma)\beta}{2C_D}}$, we have $\alpha^* < \beta^*$, which contradicts the assumption $\alpha > \beta$.

3) If $K = 0$ and $H < 0$, then $\gamma^* = 0$. Using $K = 0$ and $\alpha^* = \alpha_0 = \sqrt{\frac{(1-\gamma)\beta}{2C_D}}$, we obtain $$(\alpha^*, \beta^*, \gamma^*) = (\frac{1}{2C_A}, \frac{C_D}{2C_A^2}, 0).$$ If it satisfies $H(\alpha^*, \beta^*, \gamma^*) < 0$ and $\alpha^* > \beta^*$, it constitutes a BNE. 

4) If  $K = 0$ and $H > 0$, then $\gamma^* = \gamma_m$. Similarly, we have $$(\alpha^*, \beta^*, \gamma^*) = (\frac{1}{2C_A}, \frac{C_D}{2(1-\gamma_m)C_A^2}, \gamma_m).$$ If it fulfills $H(\alpha^*, \beta^*, \gamma^*) > 0$ and $\alpha^* > \beta^*$, it represents a BNE. 

5) If $K = 0$ and $H=0$, the solution obtained by $K=0$ and $H=0$ has measure zero, so this case is not considered.

Thus, the conclusion follows. 

\section{The proof of the Corollary \ref{lem_gamma_m} \label{app2}}

Due to Corollary \ref{U_D^*_M}, the valid solutions for \(\sigma_1\) and \(\sigma_2\) must satisfy $$1< \sigma_1 < \frac{1}{2(1-CI)},\quad \sigma_2 > \frac{1}{2(1-\gamma_m)(1-C_I)}.$$ Since \(\mU_D^*\) as a function of \(\sigma\) is increasing monotonically in both two intervals, it only remains to prove that $$\mU_D^*(\frac{1}{2(1-CI)})>\mU_D^*(\frac{1}{2(1-\gamma_m)(1-C_I)}).$$ 
Then further simplifying both sides of the inequality yields $$2C_I-1>(1-\gamma_m)(2C_I-1).$$ Since $\gamma_m<1$, this inequality obviously holds. Thus, the proof is completed.

\section{The proof of the Theorem \ref{cor5.1} \label{app3}}

From Theorem \ref{theorem_bayes} and Corollary \ref{U_D^*_B}, all BNE expressions and the defender’s benefit $\tilde{\mU}_D^*$ are independent of $\theta_1$ and $\theta_2$, and only the existence intervals depend on their ratio $\theta = \frac{\theta_1}{\theta_2}$. Since changing the probability that the insider is inadvertent does not alter the ratio $\theta$, it follows that neither the BNEs nor the defender’s benefit is affected.

\section{The proof of the Theorem \ref{the_bayes_intervel} \label{app4}}

We begin with $k_2 = M$. According to Lemma \ref{prop_1}, it is sufficient for the ratio $\sigma$ to satisfy either of the following two cases for the Bayesian strategy to yield a higher benefit than the basic one: 

1) $(\alpha^*_B, \beta^*_B)\hspace{-1mm} = \hspace{-1mm}(\frac{C_A(1-\gamma_m)^2}{2C_D^2}, \frac{1-\gamma_m}{2C_D})$, $(\alpha^*_M, \beta^*_M)\hspace{-1mm} = \hspace{-1mm}(\frac{C_A}{2C_D^2}, \frac{1}{2C_D})$.

2) $(\alpha^*_B, \beta^*_B) \hspace{-1mm}= \hspace{-0.5mm}(\frac{1}{2C_A}, \frac{C_D}{2C_A^2})$, $(\alpha^*_M, \beta^*_M)\hspace{-0.5mm} = (\frac{1}{2C_A}, \frac{C_D}{2(1-\gamma_m)C_A^2})$. 

Next, we consider the intervals of $\sigma$ corresponding to these two cases. Specifically, for the first case, according to Theorem \ref{theorem_bayes}, $(\alpha^*_B, \beta^*_B) = (\frac{C_A(1-\gamma_m)^2}{2C_D^2}, \frac{1-\gamma_m}{2C_D})$ holds if and only if 
$$ \frac{(2\theta+2)C_I-2\theta C_{AI}}{1-\gamma_m} < \sigma \leqslant \frac{1}{1-\gamma_m}.$$
Similarly, according to Corollary \ref{theorem_malicious}, $(\alpha^*_M, \beta^*_M) = (\frac{C_A}{2C_D^2}, \frac{1}{2C_D})$ holds if and only if $ \sigma \leqslant 1$. Therefore, to satisfy both conditions simultaneously, $\sigma$ must lie within the interval
\begin{equation}
\label{case1}
  \frac{(2\theta+2)C_I-2\theta C_{AI}}{1-\gamma_m} < \sigma < 1. 
\end{equation} 
For the second case, according to Theorem \ref{theorem_bayes}, $(\alpha^*_B, \beta^*_B) = (\frac{1}{2C_A}, \frac{C_D}{2C_A^2})$ holds if and only if 
$$1< \sigma < \frac{1}{(2\theta-2)C_I - 2\theta C_{AI} + 2}.$$
Similarly, according to Corollary \ref{theorem_malicious}, $(\alpha^*_M, \beta^*_M) = (\frac{1}{2C_A}, \frac{C_D}{2(1-\gamma_m)C_A^2})$ holds if and only if 
$$ \sigma>\frac{1}{2(1-C_I)(1-\gamma_m)}.$$
Therefore, to satisfy both conditions simultaneously, $\sigma$ must lie within the interval
\begin{equation}
\label{case2}
\frac{1}{2(1-C_I)(1-\gamma_m)} < \sigma < \frac{1}{(2\theta-2)C_I - 2\theta C_{AI} + 2}.    
\end{equation}
Combining the two intervals in \eqref{case1} and \eqref{case2}, we obtain $\mathcal{T}_M$.

For $k_2 = I, C$, a similar procedure applies, so the proofs are omitted for brevity.

\vfill

\end{document}